\documentclass[12pt]{article}
\usepackage{mathtools}
\usepackage{subcaption}
\usepackage{newtxmath}

\usepackage[export]{adjustbox}
\usepackage{natbib}
\usepackage{hyperref}
\usepackage{hypernat}
\usepackage{setspace}
\usepackage{tikz}

\hypersetup{breaklinks=true,
            pdfauthor={},
            pdfkeywords = {},
            pdftitle={},
            colorlinks=true,
            citecolor=blue,
            urlcolor=blue,
            linkcolor=magenta,
            pdfborder={0 0 0}}
            
\DeclareMathOperator*{\argmin}{arg\,min}
\DeclareMathOperator*{\argmax}{arg\,max}

\newcommand{\s}{\ensuremath{\mathbb{S}}}
\newcommand{\real}{\mathbb{R}}
\newcommand{\fspace}{\ensuremath{\mathcal{F}}}
\newcommand{\sign}{\mathrm{sign}}
\newcommand{\T}{\ensuremath{\mathsf{T}}}

\newcommand{\ltwo}{\ensuremath{\mathbb{L}^2}}
\newcommand{\inner}[2]{\left\langle#1,#2 \right\rangle}

\usetikzlibrary{arrows,shapes,positioning,calc,fadings,decorations.pathreplacing,backgrounds}
  \tikzset{%
    >=latex, 
    inner sep=0pt,%
    outer sep=2pt,%
    mark coordinate/.style={inner sep=0pt,outer sep=0pt, draw=blue!90,fill=blue!90,minimum size=4pt,circle},
    mark coordinate2/.style={inner sep=0pt,outer sep=0pt, draw=red!90,fill=red!90,minimum size=4pt,circle}%
  }

  \newcommand\pgfmathsinandcos[3]{%
    \pgfmathsetmacro{#1}{sin (#3)}%
    \pgfmathsetmacro{#2}{cos (#3)}%
  }
  \newcommand\LongitudePlane[3][current plane]{%
    \pgfmathsinandcos\sinEl\cosEl{#2} 
    \pgfmathsinandcos\sint\cost{#3} 
    \tikzset{#1/.style={cm={\cost,\sint*\sinEl,0,\cosEl, (0,0)}}}
  }
  \newcommand\LatitudePlane[3][current plane]{%
    \pgfmathsinandcos\sinEl\cosEl{#2} 
    \pgfmathsinandcos\sint\cost{#3} 
    \pgfmathsetmacro{\yshift}{\cosEl*\sint}
    \tikzset{#1/.style={cm={\cost,0,0,\cost*\sinEl, (0,\yshift)}}} %
  }
  \newcommand\DrawLongitudeCircle[2][1]{
    \LongitudePlane{\angEl}{#2}
    \tikzset{current plane/.prefix style={scale=#1}}
    \pgfmathsetmacro{\angVis}{atan (sin (#2)*cos (\angEl)/sin (\angEl))} %
    \draw[current plane,black!70] (\angVis:1) arc (\angVis:\angVis+180:1);
  }
  \newcommand\DrawLatitudeCircle[2][1]{
    \LatitudePlane{\angEl}{#2}
    \tikzset{current plane/.prefix style={scale=#1}}
    \pgfmathsetmacro{\sinVis}{sin (#2)/cos (#2)*sin (\angEl)/cos (\angEl)}
    \pgfmathsetmacro{\angVis}{asin (min (1,max (\sinVis,-1)))}
    \draw[current plane,black!80] (\angVis:1) arc (\angVis:-\angVis-180:1);
  }
  
\usepackage{xcolor}

\newcommand{\edit}[1]{\color{black}#1\normalcolor}

\newcommand{\blind}{1}
\title{Elastic Functional Changepoint Detection of Climate Impacts from Localized Sources}
\author{J. Derek Tucker and Drew Yarger}
\usepackage{authblk}
\author[1]{J. Derek Tucker}
\author[1]{Drew Yarger}
\affil[1]{Statistical Sciences, Sandia National Laboratories}

\date{\today}

\begin{document}

\def\spacingset#1{\renewcommand{\baselinestretch}%
{#1}\small\normalsize} \spacingset{1}

\if1\blind
{
  \baselineskip=28pt \vskip 5mm
    \begin{center} {\LARGE{\bf Elastic Functional Changepoint Detection of Climate Impacts from Localized Sources}}
    \end{center}

    \baselineskip=14pt \vskip 10mm

    \begin{center}\large
    J. Derek Tucker \footnote{\baselineskip=12pt Statistical Sciences, Sandia National Laboratories} \footnote{\baselineskip=12pt Corresponding Author - J. Derek Tucker: jdtuck@sandia.gov, Drew Yarger: anyarge@sandia.gov},
        Drew Yarger$^1$
    \end{center}
    \baselineskip=19pt \vskip 15mm \centerline{\today} \vskip 6mm
} \fi

\if0\blind
{
  \bigskip
  \bigskip
  \bigskip
  \begin{center}
    {\LARGE\bf Elastic Functional Changepoint Detection}
\end{center}
  \medskip
} \fi

\begin{center}
{\large{\bf Abstract}}
\end{center}

\edit{Detecting changepoints in functional data has become an important problem as interest in monitoring of climate phenomenon has increased, where the data is functional in nature. The observed data often contains both amplitude ($y$-axis) and phase ($x$-axis) variability.}
If not accounted for properly, \edit{true} changepoints \edit{may be un}detected, \edit{and the estimated} underlying mean \edit{change} functions will be incorrect.
In this paper, an elastic functional changepoint method is developed which properly accounts for these types of variability.  
The method can detect amplitude and phase changepoints which current methods in the literature do not, as they focus solely on the amplitude changepoint. 
This method can easily be implemented using the functions directly or \edit{can be computed via functional principal component analysis to ease the computational burden}. 
\edit{We apply the method and its non-elastic competitors to both simulated data and observed data to show its efficiency in handling data with phase variation with both amplitude and phase changepoints. 
We use the method to evaluate potential changes in stratospheric temperature due to the eruption of Mt.\ Pinatubo in the Philippines in June 1991. 
Using an epidemic changepoint model, we find evidence of a increase in stratospheric temperature during a period that contains the immediate aftermath of Mt.\ Pinatubo, with most detected changepoints occurring in the tropics as expected. 
}

\baselineskip=14pt

\par\vfill\noindent

\par\medskip\noindent
{\bf Short title}: Elastic Functional Changepoint Detection

\clearpage\pagebreak \pagenumbering{arabic}
\newpage \baselineskip=24pt

\section{Introduction}
\begin{figure}
    \centering
    \includegraphics[width = .98\textwidth]{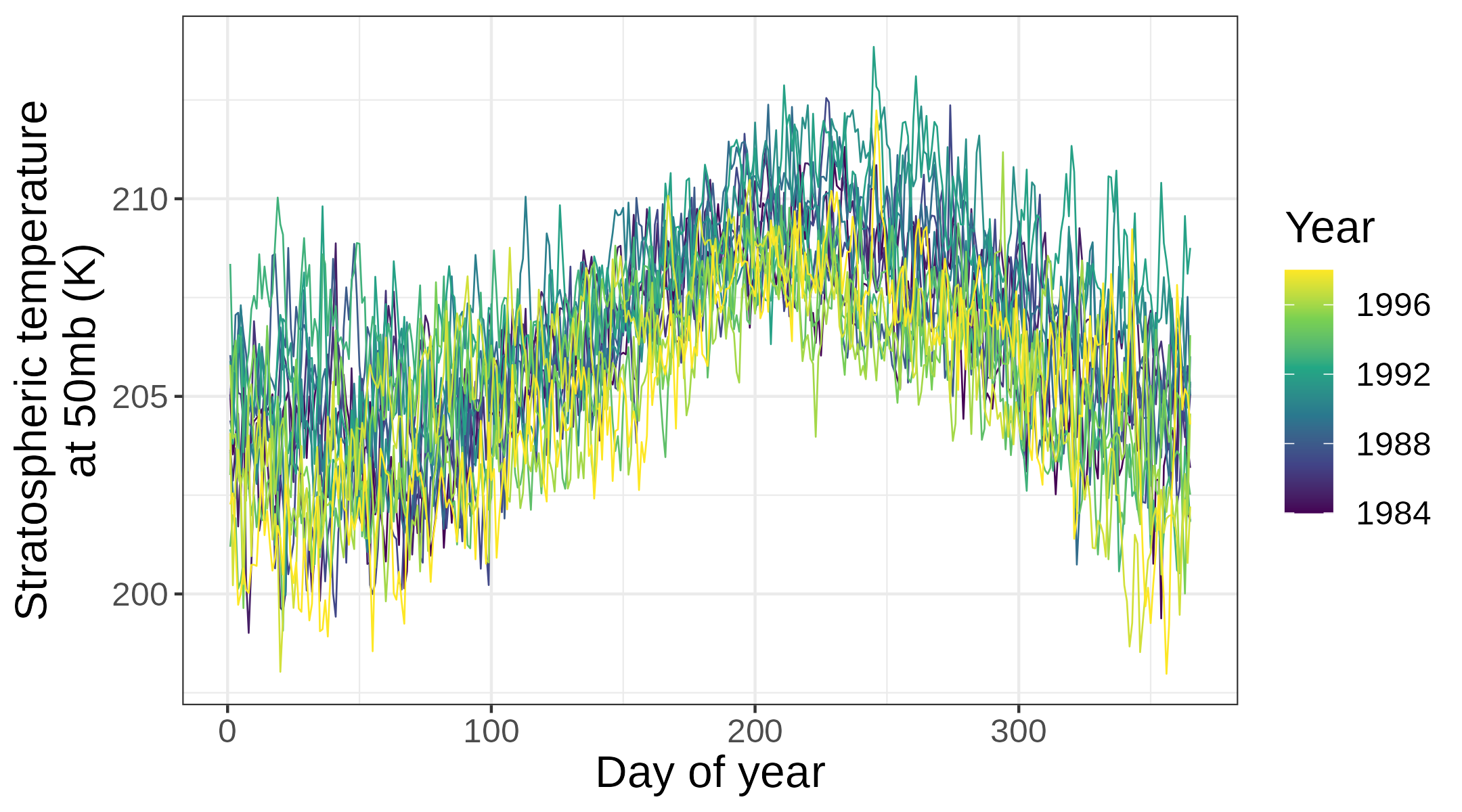}
    \caption{\edit{Daily} stratospheric temperature data presented as a functional time series at $115^\circ$E and $10^\circ$N, years 1984-1998 \edit{obtained from the Modern-Era Retrospective analysis for Research and Applications, version 2 (MERRA-2) reanalysis database} \citep{MERRA2}.}
    \label{fig:merra2_example}
\end{figure}

The statistical analysis of functional time series has become increasingly important to many scientific fields including climatology \citep{shang2011nonparametric}, finance \citep{kokoszka2012functional}, geophysics \citep{hormann2012functional}, demography \citep{hyndman2008stochastic}, manufacturing \citep{woodall2007current}, and environmental modeling \citep{fortuna2020functional}.
A functional time series is a sequence of functions (i.e., infinite dimensional objects), observed over time. 
Functional time series are analogous to univariate or multivariate time series, except that we observe a continuous function at each point in time \citep{bosq2012linear}. 
\edit{In Figure \ref{fig:merra2_example}, we plot daily stratospheric temperatures at one location as a functional time series, one of the motivating data sets for this work.
In each year, we observe a function that describes the temperature over the course of that year. Although not obvious in Figure \ref{fig:merra2_example}, stratospheric temperature trends experienced a significant change from normal following the 1991 Mt Pinatubo eruption.}

Similar to univariate and multivariate time series, a functional time series can experience abrupt changes in its data-generating mechanism. 
These abrupt changes are known as changepoints and can complicate further analysis by invalidating stationarity assumptions. 
They also can lead to interesting analysis by determining when data has patterns which are not homogeneous over time, or when a forcing has caused the data-generation mechanism to change.
\edit{We consider the problem of detecting changepoints after observing an entire sequence of functions rather than considering a ``streaming'' or ``real-time'' changepoint detection problem.}
\edit{That is, we are interested in evaluating (1) whether a change occurred in the functional time series or not and (2) if a change was detected, when the change occurred and how the data-generating mechanism changed. }
\edit{This is particularly useful for changepoints that can traverse multiple years, which is the case for our motivating problem.}

Current methods in the Functional Data Analysis (FDA) literature on changepoint detection have been solely focused on cross-sectional analysis, \edit{or amplitude variability only, and thus} completely ignore phase variability that \edit{often} exist in the data, \edit{especially climate data}. 
In \cite{berkes:2009}, a Cumulative Sum (CUSUM) test was proposed for independent functional data, which was further studied in \cite{aue2009estimation}, where its asymptotic properties were developed. 
This test was then extended to weakly dependent functional data by \cite{hormann2010weakly}. \cite{zhang2011testing} introduced a test for changes in the mean of weakly dependent functional data using self-normalization to alleviate the use of asymptotic control. 
\cite{sharipov2016sequential} similarly developed a sequential block bootstrap procedure for these methods.
More recently, \cite{gromenko:2017} considered changes in spatially-correlated functional data, and \cite{aue:2018} proposed a fully functional method for finding a change in the mean without losing information due to functional principal component analysis (fPCA) truncation.

\begin{figure}[htbp]
	\centering
	\begin{subfigure}{0.48\textwidth}
	\centering
	\includegraphics[width=\textwidth]{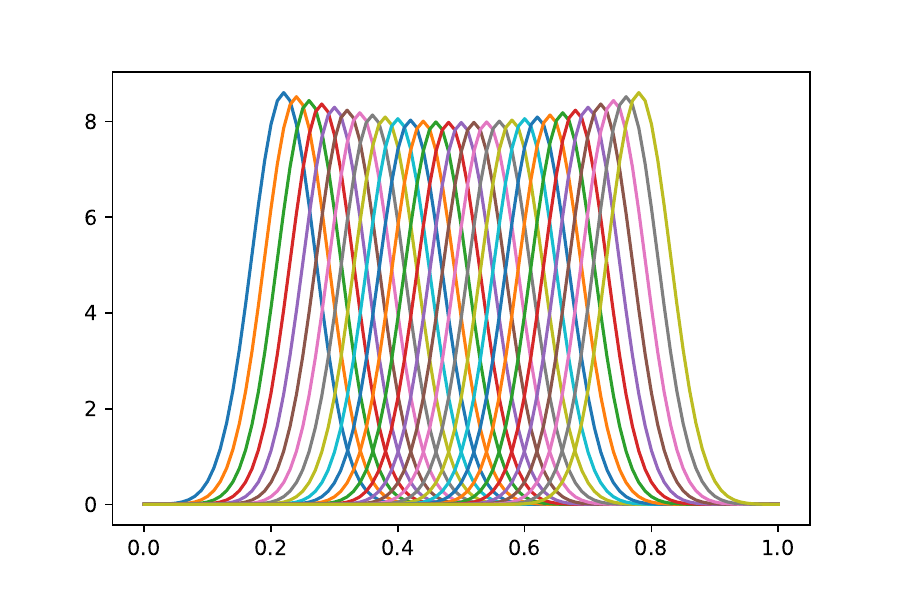}		
	\label{fig:origtoy}
	\caption{Original Data}	
	\end{subfigure}
	\hfill
	\begin{subfigure}{0.48\textwidth}
	\centering
	\includegraphics [width=\textwidth]{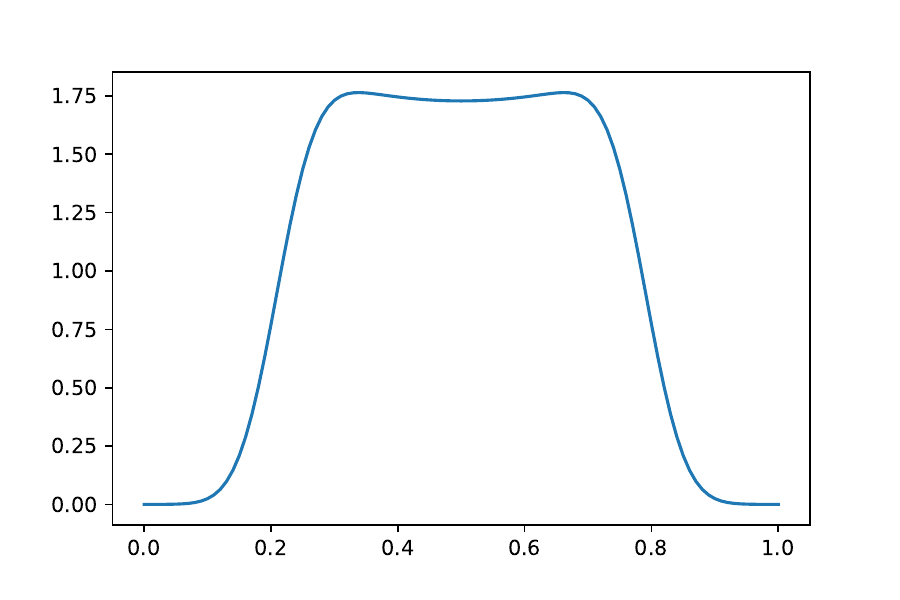}	
	\label{fig:origtoy_mean}
	\caption{Original Mean}	
	\end{subfigure}
	\hfill
	\begin{subfigure}{0.48\textwidth}
	\centering
	\includegraphics [width=\textwidth]{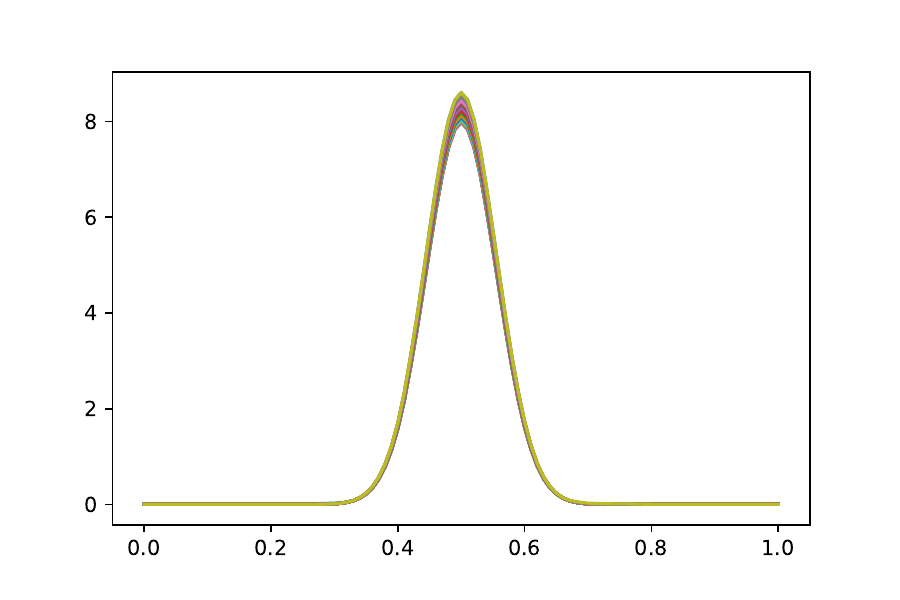}	
	\label{fig:alignedtoy}
	\caption{Aligned Data}	
	\end{subfigure}
	\hfill
	\begin{subfigure}{0.48\textwidth}
	\centering
	\includegraphics[width=\textwidth]{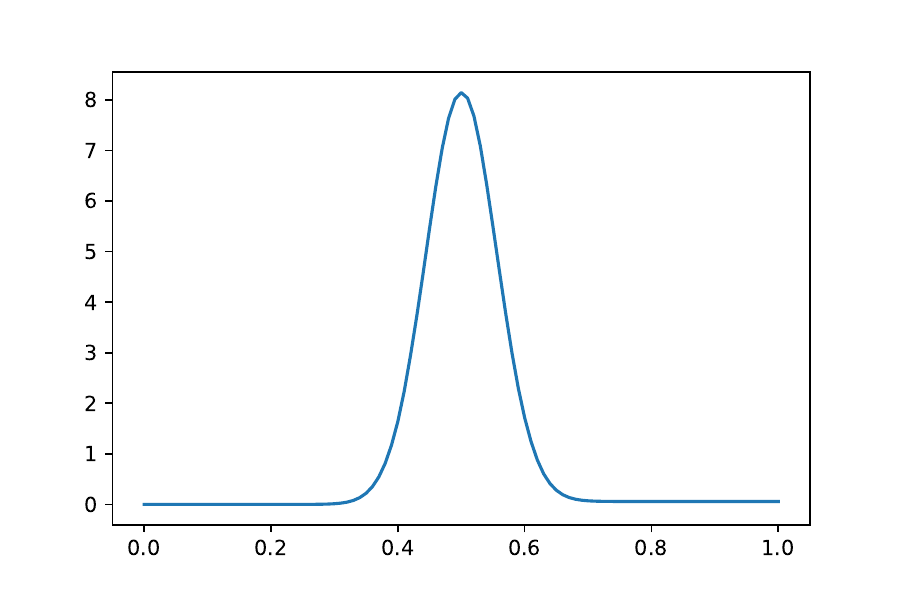}
	\caption{Karcher Mean}	
	\label{fig:alignedtoy_mean}
	\end{subfigure}

	\caption{Example of using proper mean calculation on functional data. Panel (a) presents a set of functional data that contains a large amount of phase variability. Panel (b) shows the computed mean if the phase variability is ignored. 
	Panel (c) shows the alignment of the data using the method in \cite{tucker:2013} and the resulting Karcher Mean in Panel (d).}
    \label{fig:toy_data}
\end{figure}

\edit{Phase variability could be just a nuisance parameter or actually contain information. In the latter case, if phase variability is ignored, the distance metric between functions will be degenerate and the estimated underlying mean function will not be representative of the data-generating mechanism (\cite{tucker:2013} and \cite{srivastava2016functional}).} 
An example of this is shown in Figure~\ref{fig:toy_data}.
In Panel (b), it is \edit{clear} that the mean is not representative of the underlying data-generating mechanism demonstrated in Panel (a). 
After alignment shown in panel (c), the mean in Panel (d) \edit{accounts for phase variability and} is representative of the underlying data-generating mechanism.

In this paper we present a functional changepoint method that accounts for the amplitude and phase variability. 
We demonstrate \edit{two different approaches for changepoint detection in this setting:}: 1) by utilizing the appropriate fPCA \citep{tucker:2013,lee:2017} or 2) by utilizing a fully functional approach similar to \cite{aue:2018} where the mean is computed using the Karcher mean under the Fisher-Rao metric \citep[][and described later]{srivastava2016functional}. 
This Fisher-Rao metric is a proper distance and is \textit{elastic} (that is, it is invariant to random warping). 
Through this \edit{metric} we separate out the amplitude and phase variability for subsequent changepoint analysis. 
\edit{This allows for the identification of} changepoints where the underlying mean changes in amplitude, phase, or a combination of both. 
\edit{We evaluate performance in two simulation studies, showing that the proposed test statistics properly identify changes and perform appropriately under the null hypothesis. }

\edit{A motivation for developing this methodology is to introduce methodology to evaluate changes the temporal pattern of stratospheric temperatures after the eruption of Mt.\ Pinatubo in June 1991, as presented in Figure \ref{fig:merra2_example}. 
Using the years 1984-1998, we aim to estimate a change in this sequence of functional data during these $n=15$ years. 
There is evidence that the eruption of Mt.\ Pinatubo and its injection of aerosols into the atmosphere increased temperatures in the stratosphere, leading to many downstream shocks to the climate system  \citep{robock2003introduction}. 
\edit{Climate data is known to exhibit strong phase variability in the form of} transient or cyclic variability from year to year.
In our analysis of stratospheric temperature, we find that controlling for phase variability isolates changes in the amplitude while filtering out transient weather variability from different years. 
When identifying changes in climate data, controlling for seasonality is an important aspect of many analyses; see \cite{reeves_2007} and more recently \cite{lund_2022} for an application, overview, and common challenges for detecting changes in general (non-functional) climate time series data.
The data in Figure \ref{fig:merra2_example} demonstrates substantial seasonality in stratospheric temperature.
By describing seasonality using smooth functions, the proposed model handles general seasonality characteristics of the data and retains the high-resolution nature of daily-averaged data. 
}

The remainder of the paper is organized as follows: 
Section \ref{sec:efchange} presents the elastic amplitude and phase changepoint methods, and outlines the implementation using fully functional or functional principal component analysis.
Section \ref{sec:results} provides analysis of the methodologies on simulated and \edit{multiple climate variables from the MERRA-2 data base.}
Finally, concluding remarks and future work are presented in Section \ref{sec:conclusion}.

\section{Elastic Functional Changepoint Analysis} \label{sec:efchange}

\subsection{The Functional Changepoint Problem}
Assume we have real-valued functions $f_1,\dots,f_n$ that are absolutely continuous on the interval $[0,1]$. 
Using this interval is not a restriction and can be easily extended to any interval.
The standard changepoint problem \citep{berkes:2009,aue:2018} assumes the data is from the following model:
\[
f_i = \mu + \delta \vmathbb{1}(i > k^*) + \epsilon_i,
\]
\edit{where $\lVert \mu \rVert^2 < \infty$, $\lVert \delta \rVert^2 < \infty$, and $\mathbb{E}\lVert \epsilon_i \rVert^2 < \infty$ and $\lVert \cdot \rVert$ is the standard $\mathbb{L}^2$ functional norm.}
The point $k^*$ labels the time of the unknown mean change and $\epsilon_i$ is a zero mean random error term with may be i.i.d.\ or potentially weakly dependent as in \cite{aue:2018}.
The changepoint {\color{red} detection} problem then becomes a hypothesis testing problem of 
\[H_0: \delta = 0~~\text{vs}~~H_A: \delta\neq 0.\]
\cite{aue:2018} considers the test statistic of $T_n = \max_{1 \leq  k \leq n} \lVert S_{n,k}\rVert^2$, where \begin{align*}
    S_{n,k} = \frac{1}{\sqrt{n}} \left( k\mu_k -  k \mu_n\right),
\end{align*}$\mu_k = k^{-1} \sum_{i=1}^k f_i$.
\cite{aue:2018} demonstrate asymptotic properties of this test statistic. 
{\color{red} In addition to determining whether there is evidence of a change or not (the detection problem), \cite{aue:2018} also give an estimate of $k^*$ using $\hat{k}^*  =\argmax_{1\leq k\leq n}\lVert S_{n,k}\rVert^2$ (the estimation problem). }

{\color{red} In addition to the fully-functional statistic, }\cite{berkes:2009} and subsequent work propose a functional principal component-based test statistic that includes dimension reduction. 
{\color{red}This test statistic evaluates evidence of a change in the mean of the functional data, and \cite{berkes:2009} also propose an estimate of $k^*$ which was later adopted in \cite{aue2009estimation}.}
These two approaches will motivate our test statistics for functional data with phase and amplitude variability. 

Since the test statistics of \cite{aue:2018} and \cite{berkes:2009} are based \edit{on} functional statistics that do not use functional alignment, this cross-sectional approach assumes there is no phase variability and that there is only a random amplitude term. 
We know from \cite{tucker:2013} and \cite{srivastava2016functional} that functional data contains two types of variation, amplitude and phase, and both are important to consider when evaluating functional data for statistical analysis and modeling. 

\subsection{The Elastic Functional Changepoint Problem}

\edit{Before introducing the changepoint problem for functional data with amplitude and phase variation, we review key concepts regarding the separation of amplitude and phase variation in functional data. 
First, we will introduce the group of warping functions that represent actions on the phase of functions. }
Let $\gamma$ denote a warping function that is from the following group $\Gamma$, the phase space of the unit interval $[0, 1]$ defined as 
\[
\Gamma = \{\gamma : [0, 1] \mapsto [0, 1] \ | \ \gamma(0) = 0 \text{, } \gamma(1) = 1 \text{, } \gamma \text{ is a diffeomorphism}\}.
\]
This diffeomorphic constraint gives rise to the notion of elasticity because the elements of $\Gamma$, i.e.\ phase functions, can only smoothly stretch and contract portions of the unit interval so that it maps back to itself.
It should also be noted, that defining on the interval $[0,1]$ is not a restriction and all the following construction can be extended to any arbitrary interval. 
\edit{The set $\Gamma$ is a group under the composition operation $\circ$ and has identity element $\gamma(t) = t$.}
Additionally, we will use the Fisher-Rao metric and the square-root velocity function (SRVF) for its attractive properties (see \cite{srivastava2016functional} for properties and appropriate proofs). 
The square-root velocity function \edit{(SRVF)} is defined as:
\[q = \sign(\dot{f})\sqrt{|\dot{f}|}\]
where $\dot{f}$ is the time derivative of the function $f$. 
\edit{The Fisher-Rao metric and for functions $f_1$ and $f_2$ with corresponding SRVFs $q_1$ and $q_2$,is defined as 
\begin{align*}
    d_{FR}(f_1, f_2) = \lVert q_1 - q_2 \rVert^2
\end{align*}
which is the $\mathbb{L}^2$ distance between the SRVFs.}
\edit{Define $(q,\gamma) = q\circ\gamma\sqrt{\dot{\gamma}}$ to represent the effect of a warping function to a square-root velocity function. That is, if the SRVF of $f$ is $q$, then the SRVF of $f\circ \gamma$ is $(q, \gamma)$. }
\edit{An attractive property of the Fisher-Rao distance is its invariance to random warping, so that for two SRVFs $q_1$ and $q_2$ and a warping function $\gamma$, $\lVert (q_1, \gamma) - (q_2, \gamma)\rVert^2 = \lVert q_1 - q_2\rVert^2$.}

\edit{To consider a changepoint problem for the amplitude of functional data that has phase variability,} we propose the model
\[q_i = \left( (\mu_q + \delta_q \vmathbb{1}(i > k^*) + \epsilon_i), \gamma_i^{-1}\right).
\]
\edit{Here $\mu_q$ is the mean of the square-root velocity functions before a change $k^*$, $\delta$ is its change, $\{\epsilon_i\}_{i=1}^n$ are random, zero-mean functions representing amplitude variability, and $\{\gamma_i\}_{i=1}^n$ are random warping functions that represent phase variability.  }
Here, we now aim to detect a changepoint in the sequence of SRVF functions while controlling for phase variability. 
We can also view the data as 
\[(q_i,\gamma_i) = \mu_q + \delta_q \vmathbb{1}(i > k^*) + \epsilon_i.\]
For the changepoint detection problem, we wish to test the hypothesis 
\[H_0: \delta_q = 0~~\text{vs}~~H_A: \delta_q \neq 0.\]
We are also interested in the estimation problem of $k^*$ and the mean elements $\mu_q$ and $\delta_q$. 

\edit{We next turn to a different model that hypothesizes a change in the phase of the functional data.
Now let \begin{align*}
    (q_i, \gamma_i) = \mu_q + \epsilon_i.
\end{align*}
In this setting, we consider a model for $\gamma_i$ as \[\gamma_i =\begin{cases}
\eta_i(\mu_{\gamma}) & \textrm{if } i \leq k^* \\
\eta_i(\nu_{\gamma}) & \textrm{if } i > k^*
\end{cases}\]where $\eta_i(\mu)$ denotes a random element of $\Gamma$ with mean $\mu \in \Gamma$.
Therefore, $\mu_{\gamma}\in \Gamma$ and $\nu_{\gamma}\in \Gamma$ are the mean warping functions before and after the change, respectively. 
We then aim to test the hypothesis
\[H_0: \mu_{\gamma}= \nu_{\gamma}~~\text{vs}~~H_A: \mu_{\gamma}\neq \nu_{\gamma}.\]
Again, we also consider the estimation problem of $k^*$ as well as estimating $\mu_q$, $\mu_{\gamma}$, and $\nu_{\gamma}$.}

In the next two subsections, we propose a test procedure for changepoints in the amplitude mean and phase mean of functional data, respectively. 
\edit{While we have separated the amplitude and phase changepoint detection problems, there may be changes in both the amplitude and phase of the functional data. 
In this setting, by properly disentangling these two different sources of variability, one can evaluate changes without the sources of variability potentially impeding each other. If the practitioner has no prior knowledge on what type of changepoint may exist in the data, one may test for both an amplitude and a phase change and use a Bonferroni correction to the resulting p-values. 
However, in many cases, the type of change may be known a priori. 
For example, we are primarily interested in an amplitude change in our data application.}

\subsection{Amplitude changepoint}
\label{sec:ampchange}
From \cite{srivastava2016functional} the amplitude distance (which is a proper distance) between two functions is defined as
\[d_a(f_1,f_2)=\inf_{\gamma\in\Gamma} \|q_1-(q_2,\gamma)\|.\]
\edit{Notice that $d_a(f_1, f_2)$ is the infimum of the Fisher-Rao distance between the square-root velocity function $q_1$ and a warped version of the SRVF $q_2$. }
From a proper distance we can define the Karcher mean of functions as 
\[\mu_q^k = \argmin_{q} \sum_{i=1}^k d_a(q,q_i)^2.\]
\edit{It should be noted that since the solution to this optimization problem is invariant to random warping and technically produces an orbit (i.e., a set of solutions). 
We pick the element of the orbit similar to \cite{tucker:2013} where the optimizer is the one corresponding to where the mean of ${\gamma_i}$ is $\gamma_{id}$.
Notice that the alignment process and the computation of a Karcher mean do not take into account potential dependence of elements in the functional time series. }

To determine if an amplitude changepoint has occurred, we can define a test statistic using the (scaled) functional cumulative sum. 
This is similar to the approach used in \cite{aue:2018}.
The test statistic is defined as
\begin{equation}
	S_{n,k} = \frac{1}{\sqrt{n}}\left(k\mu_q^k-k\mu_q^n\right)\label{eq:ff_amp}
\end{equation}
where $\mu_q^n$ is the Karcher mean of the entire data set and for a specific value of $k$ we can compute the Karcher mean $\mu_{q}^k$.
We can also define a test statistic using the $\ltwo$ norm to also test for differences in means which is similar to the test statistic proposed in \cite{gromenko:2017}.
The test statistic is defined as
\begin{equation}
	\Lambda_2 = \frac{1}{n}\sum_{k=1}^n d_a(k\mu_q^k , k\mu_q^n )^2.
\end{equation}

One can \edit{use} the max-point structural break detector
\[T_n = \max_{1\leq k \leq n} \|S_{n,k}\|^2\]
to test the hypothesis $H_0$ versus $H_A$.
\edit{One may} also estimate the changepoint \edit{$\hat{k}^*$} where
\[\hat{k}^* = \min\{k:\|S_{n,k}\|= \max_{1\leq k \leq n} \|S_{n,k}\|\}.\]
For the asymptotic properties of this statistic the reader is referred to \cite{aue:2018}. 
Under this model $T_n$ is distributed according to
\[T_n \xrightarrow{d} \sup_{0\leq x\leq 1} \sum_{l=1}^\infty \lambda_l B_l^2(x)\]
where $B_l$ are i.i.d.\ standard Brownian bridges defined on $[0,1]$ and $\{\lambda_l\}$ are the eigenvalues on the long-run covariance kernel of the error sequence $\epsilon_i$. 
\edit{For this theorem, we assume that the errors $\epsilon_i(u)$ are $L^p-m$ approximable, and that again $\lVert \mu \rVert^2 < \infty$, $\lVert \delta \rVert^2 < \infty$, and $\mathbb{E}\lVert \epsilon_i \rVert^2 < \infty$; see \cite{aue:2018} for the definition of $L^p-m$ approximable, which allows for some dependence between elements in the functional time series.}
The estimation of $\{\lambda_\ell\}$ can then be done using the estimator in Section 3.1 of \cite{aue:2018}.
Computation of critical values of $T_n$ can be done by the simulation of independent Brownian bridges and truncating the infinite sum; see Section 3.2 of \cite{aue:2018} for more details on this computation. 

\edit{The test statistic} $T_n$ evaluates potential mean shape changes \edit{for functional data} under phase variability. 
We can also easily implement this using functional principal component analysis (fPCA) as in \cite{berkes:2009}.
In order to properly account for the variability, we can use the vertical
fPCA and horizontal fPCA presented in \cite{tucker:2013}.
These fPCA methods account for the variability by first separating the
phase and amplitude and then performing the fPCA on the spaces
separately. 
Using these methods one can construct the changepoint test in the 
amplitude space using the SRVF, \(q\), or
specifically the aligned SRVF, \(\tilde{q}\), and in the phase space using the warping
functions, \(\gamma\).

We can use the \textit{vertical} fPCA from \cite{tucker:2013} to test for an amplitude change. We will first give a short review of this fPCA method below\edit{, and more details can be found in \cite{tucker:2013}.}

\subsubsection{Vertical Functional Principal
Components}
Let \(f_1,\cdots,f_n\) be a given set of functions, and
\(q_1,\cdots,q_n\) be the corresponding SRVFs, \({\mu}_q\) be their
Karcher Mean, and let \(\tilde{q}_i\)s be the corresponding aligned
SRVFs using Algorithm 1 from \cite{tucker:2013}. In
performing vertical fPCA, one also needs to include the variability
associated with the initial values, i.e., \(\{f_i(0)\}\), of the given
functions. 
Since representing functions by their SRVFs ignores
vertical translation, this variable is treated separately. That is, a
functional variable \(f\) is analyzed using the pair \((q, f(0))\)
rather than just \(q\). This way, the mapping from the function space
\(\fspace\) to \(\ltwo \times \real\) is a bijection. In practice, where
\(q\) is represented using a finite partition of \([0,1]\), say with
cardinality \(T\), the combined vector \(h_i = [q_i~~f_i(0)]\) simply
has dimension \((T+1)\) for fPCA. We can define a sample covariance
operator for the aligned combined vector
\(\tilde{h}_i = [\tilde{q}_i~~f_i(0)]\) as 
\[K_h = \frac{1}{n-1}\sum_{i=1}^n (\tilde{h}_i - \mu_h)(\tilde{h}_i - \mu_h)^\T \in \mathbb{R}^{(T+1) \times (T+1)}\ ,\]
where \(\mu_h = [\mu_q~~ \bar{f}(0)]\). 
Taking the SVD,
\(K_h=U_h\Lambda_h V_h^\T\), we can calculate the directions of principal
variability in the given SRVFs using the first \(p\leq n\) columns of
\(U_h\) and can then \edit{convert} back to the function space \(\mathcal{F}\),
via integration, \edit{to find} the principal components of the original
functional data. Moreover, we can calculate the observed principal
scores as \(\eta_{i, l} = \inner{\tilde{h}_i}{U_{h,l}}\).

The test statistic would then follow directly from \cite{berkes:2009} with test statistic
\begin{equation}
	\label{eq:tn}	
	T_N(x) = \frac{1}{N}\sum_{l=1}^d\lambda_l^{-1}\left(\sum_{1\leq i \leq Nx} \eta_{i,l}-x\sum_{1\leq i\leq N} \eta_{i,l}\right)^2
\end{equation}
where $\lambda$ are the eigenvalues from $\Lambda_h$ and $\eta$ are the scores from the fPCA.
The test statistic is distributed as
\[T_N(x) \xrightarrow{d} \sum_{1\leq l \leq d} B_l^2(x),~~0\leq x \leq 1\]
where $B_l(\cdot)$ are independent Brownian Bridges on $[0,1]$. 

\subsection{Phase changepoint}
\label{sec:phasechange}
From the calculation of the Karcher Mean, $\mu_q^k$, we have the set of warping functions $\{\gamma_i\}^k$.
To perform a changepoint test on the space of warping functions directly is difficult, as $\Gamma$ is a infinite-dimensional nonlinear manifold and computation of the mean is not straightforward due to the unknown geometry.
To facilitate computing a mean, we can simplify the geometry of $\Gamma$.

\subsubsection{Simplifying Geometry of $\Gamma$}\label{sec:phase_geometry}
The space of warping functions, \(\Gamma\), is an infinite-dimensional
nonlinear manifold and therefore cannot be treated as a standard
Hilbert space. To overcome this problem, we will use tools from
differential geometry to perform statistical analyses and model the
warping functions. The following framework was previously used in
various settings including (1) modeling re-parameterizations of curves
\citep{srivastava-jermyn-PAMI:09} and (2) putting prior distributions on
warping functions \citep{Kurtek17,tucker:21}, and many others. It is also very
closely related to the square-root representation of probability density
functions introduced by \cite{bhattacharya-43}.

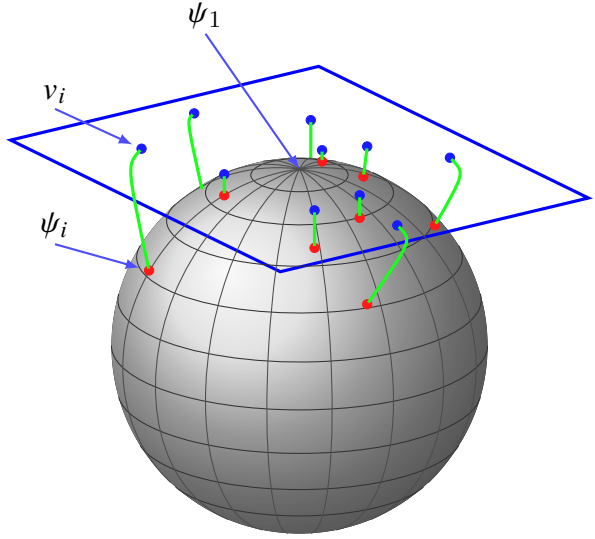
\begin{figure}[!t]
\begin{center}
  \begin{tikzpicture}
  	\def\R{2.5} 
    \def\angEl{20} 
	\def\angAz{-55} 
	\def\angPhiOne{67} 
	\pgfmathsetmacro\H{\R*cos(\angEl)} 
	\tikzset{xyplane/.estyle={cm={cos(\angAz),sin(\angAz)*sin(\angEl),-sin(\angAz),
	                              cos(\angAz)*sin(\angEl),(0,\H)}}}
	\tikzset{dot/.style={circle,draw=red!90,fill=red!90,minimum size=3.5pt}}
	\tikzset{dot2/.style={circle,draw=blue!90,fill=blue!90,minimum size=3.5pt}}
	\LongitudePlane[xzplane]{\angEl}{\angAz}
	\LongitudePlane[pzplane]{\angEl}{\angPhiOne}
	\LatitudePlane[equator]{\angEl}{0}
	\node at (.15,\R-1) [dot] (J) {};
	\node at (.15,\R+.5) [dot2] (10) {};
	\draw[green!90,line width=1.2pt, shorten <=-3pt,shorten >=-3pt] (J) -- (10);
	\fill[ball color=gray!30] (0,0) circle (\R);
	\draw[xyplane,blue,line width=1.2pt] (-1.25*\R,-1*\R) rectangle (1.25*\R,1*\R);
	\coordinate (O) at (0,0);
	\coordinate (N) at (0,\H);
	\coordinate[label=above:$\psi_1$] (N1) at (-1.25,1.8+\H);
	\coordinate[label=above:$v_i$] (v) at (-3.25,.8+\H);
	\coordinate[label=above:$\psi_i$] (N2) at (-3.25,\H-1);
	\DrawLatitudeCircle[\R]{0};
	\DrawLatitudeCircle[\R]{15};
	\DrawLatitudeCircle[\R]{30};
	\DrawLatitudeCircle[\R]{45};
	\DrawLatitudeCircle[\R]{60};
	\DrawLatitudeCircle[\R]{75};
	\DrawLatitudeCircle[\R]{89};
	\DrawLatitudeCircle[\R]{-15};
	\DrawLatitudeCircle[\R]{-30};
	\DrawLatitudeCircle[\R]{-45};
	\DrawLongitudeCircle[\R]{160};
	\DrawLongitudeCircle[\R]{140};
	\DrawLongitudeCircle[\R]{120};
	\DrawLongitudeCircle[\R]{100};
	\DrawLongitudeCircle[\R]{80};
	\DrawLongitudeCircle[\R]{60};
	\DrawLongitudeCircle[\R]{40};
	\DrawLongitudeCircle[\R]{20};
	\node at (2,\R) [dot2] (1) {};
	\node at (.8,\R-.5) [dot2] (2) {};
	\node at (1.3,1.6) [dot2] (3) {};
	\node at (-1.4,\R+.59) [dot2] (4) {};
	\node at (-2.1,\R+.12) [dot2] (5) {};
	\node at (-1,\R-.22) [dot2] (6) {};
	\node at (.9,\R+.15) [dot2] (7) {};
	\node at (.3,\R+.1) [dot2] (8) {};
	\node at (.2,1.8) [dot2] (9) {};
	\node at (.3,\R-.05) [dot] (A) {};
	\node at (.9,.55) [dot] (B) {};
	\node at (1.8,1.6) [dot] (C) {};
	\node at (-1.3,2.1) (D) {};
	\node at (-2,1) [dot] (E) {};
	\node at (.8,1.7) [dot] (F) {};
	\node at (-1,2) [dot] (G) {};
	\node at (.85,\R-.25) [dot] (H) {};
	\node at (.2,1.3) [dot] (I) {};
	\draw[-latex,blue!70,thick] (N1) -- (N);
	\draw[-latex,blue!70,thick] (v) -- (5);
	\draw[-latex,blue!70,thick] (N2) -- (E);
	\draw[green!90,line width=1.2pt, shorten <=-3pt,shorten >=-3pt] (B) .. controls (1.5,1.4) .. (3);
	\draw[green!90,line width=1.2pt, shorten <=-3pt,shorten >=-3pt] (C) .. controls (\R-.3,2.3) .. (1);
	\draw[green!90,line width=1.2pt, shorten <=-3pt,shorten >=-3pt] (D) .. controls (-1.5,\R+.5) .. (4);
	\draw[green!90,line width=1.2pt, shorten <=-3pt,shorten >=-3pt] (E) .. controls (-2.3,\R-.1) .. (5);
	\draw[green!90,line width=1.2pt] (A) -- (8);
	\draw[green!90,line width=1.2pt, shorten <=-3pt,shorten >=-3pt] (F) -- (2);
	\draw[green!90,line width=1.2pt, shorten <=-3pt,shorten >=-3pt] (G) -- (6);
	\draw[green!90,line width=1.2pt, shorten <=-3pt,shorten >=-3pt] (H) -- (7);
	\draw[green!90,line width=1.2pt, shorten <=-3pt,shorten >=-3pt] (I) -- (9);
\end{tikzpicture}
  \caption{Depiction of the SRVF space of warping functions as a sphere and a tangent space at $\psi_{1}$.}
  \label{fig:sphere-map}
\end{center}
\end{figure}

We represent an element \(\gamma \in \Gamma\) by the square-root of its
derivative \(\psi = \sqrt{\dot{\gamma}}\). Note that this is the same as
the SRVF defined earlier, and it takes this form since
\(\dot{\gamma} > 0\). The identity \(\gamma_{id}\) maps to a constant
function with value \(\psi_{id}(t) = 1\). Since \(\gamma(0) = 0\), the
mapping from \(\gamma\) to \(\psi\) is a bijection and one can
reconstruct \(\gamma\) from \(\psi\) using
\(\gamma(t) = \int_0^t \psi(s)^2 ds\). An important advantage of this
transformation is that since
\(\| \psi\|^2 = \int_0^1 \psi(t)^2 dt = \int_0^1 \dot{\gamma}(t) dt = \gamma(1) - \gamma(0) = 1\),
the set of all such \(\psi\)s is the positive orthant of the unit Hilbert
sphere in \(\ltwo\): \(\Psi=\s_{\infty}^+\). In other words, the square-root representation simplifies
the complicated geometry of \(\Gamma\) to a (subset of a) unit sphere. The distance
between any two warping functions, i.e., the phase distance, is exactly
the arc-length between their corresponding SRVFs on \(\Psi\):
\begin{equation}
  d_{p}(\gamma_1, \gamma_2) = d_{\psi}(\psi_1, \psi_2) \equiv \cos^{-1}\left(\int_0^1 \psi_1(t) \psi_2(t) dt \right)\ .
\end{equation}
Figure \ref{fig:sphere-map} depicts the SRVF space
of warping functions as a unit sphere \citep{tucker-2014}.

While the geometry of \(\Psi\subset\s_{\infty}\) is more tractable, it
is still a nonlinear manifold and computing standard statistics remains
difficult. Instead, we use a tangent (vector) space at a certain fixed
point for further analysis. The tangent space at any point
\(\psi \in \Psi\) is given by:
\(T_{\psi}(\Psi) = \{v \in \ltwo| \int_0^1 v(t) \psi(t) dt = 0\}\). To
map between the representation space \(\Psi\) and tangent spaces, one
requires the exponential and inverse-exponential mappings. The
exponential map at a point \(\psi\in\Psi\) denoted by
\(\exp_\psi : T_{\psi}(\Psi) \mapsto \Psi\), is defined as

\begin{equation}
  \exp_\psi(v) = \cos(\|v\|)\psi+\sin(\|v\|)\frac{v}{\|v\|},
  \end{equation} \noindent where \(v\in T_{\psi}(\Psi)\). Thus,
\(\exp_\psi(v)\) maps points from the tangent space at \(\psi\) to the
representation space \(\Psi\). Similarly, the inverse-exponential map,
denoted by \(\exp_{\psi}^{-1} : \Psi \mapsto T_{\psi}(\Psi)\), is
defined as

\begin{equation}
  \exp_{\psi}^{-1}(\psi_1) = \frac{\theta}{\sin(\theta)}(\psi_1-\cos(\theta)\psi),
  \end{equation} \noindent where \(\theta = d_{p}(\gamma_1, \gamma)\).
This mapping takes points from the representation space to the tangent
space at \(\psi\).

The tangent space representation \(v\) is sometimes referred to as a
\emph{shooting vector}, as depicted in Figure \ref{fig:sphere-map}. The
remaining question is which tangent space should be used to represent
the warping functions. A sensible point on \(\Psi\) to define the
tangent space is at the sample Karcher mean \(\hat{\mu}_{\psi}\)
(corresponding to \(\hat{\mu}_{\gamma}\)) of the given warping
functions. We can compute the mean of the warping functions using Algorithm 2 from \cite{tucker:2013}.

We then can perform a test if we have a changepoint in the phase space (on the tangent space of $\s_\infty$) using the test statistic
\begin{equation}
	S_{n,k} = \frac{1}{\sqrt{n}}\left(k\bar{v}^k - k \bar{v}^n\right),
\end{equation}
which is similar as defined previously, where $\bar{v}^k$ is the shooting vector of the Karcher mean of the warping functions, $\hat{\mu}_\psi$, computed using the first $k$ warping functions.
We can also define the test using the $\ltwo$ norm similar to \cite{gromenko:2017} as
\[\Lambda_2 = \frac{1}{n}\sum_{k=1}^n \| k\bar{v}^k - k\bar{v}^n\|^2.\]
Under this test we are testing if there is a phase shift or changepoint.
The definition of $T_n$, its distribution, and finding \edit{$\hat{k}^*$} are the same as the amplitude changepoint in Section~\ref{sec:ampchange}.
As mentioned above we can also implement this using \textit{horizontal} fPCA from \cite{tucker:2013}. 

\subsubsection{Horizontal Functional Principal
Components}
To perform horizontal fPCA we will use the tangent space at
\(\mu_{\psi}\) \citep{srivastava2016functional} to perform analysis, where \(\mu_{\psi}\) is the mean of
the transformed warping functions.
In this tangent space we can define a sample covariance function:
\[K_\psi = \frac{1}{n-1} \sum_{i=1}^n v_i v_i^\T \in \real^{T\times T}.\]
The singular value decomposition (SVD) of
\(K_{\psi} = U_{\psi} \Lambda_{\psi} V_{\psi}^\T\) provides the estimated
principal components of \(\{ \psi_i\}\): the principal directions
\(U_{\psi,l}\) and the observed scores
\(\eta_{i,l} = \inner{v_i}{U_{\psi,l}}\). This analysis on \(\s_{\infty}\) is similar
to the ideas presented in \cite{srivastava-joshi-etal:05} although one
can also use the idea of principal nested sphere for this analysis
\citep{jung-dryden-marron:2012}. 

We can then use the test defined in Equation \ref{eq:tn} to perform the changepoint test. 
The distribution to find critical point and confidence interval apply directly.

\section{Application}
\label{sec:results}
\defcitealias{era5}{ECMWF}
To demonstrate our proposed method, we will compare our elastic method to the cross-section changepoint methods of \cite{aue2009estimation} and \cite{aue:2018}.
First, we will assess our method's performance on simulated data where an amplitude changepoint exists as well as simulated data where a phase changepoint exists. 
We then apply our method to a global temperature observations dataset based on the MERRA-2 reanalysis product \citep{MERRA2} during the time period (1984-1998) that contains the eruption of Mt.\ Pinatubo in the Philippines in June 1991.

\subsection{Simulation Data}
\label{sec:simulation}

\begin{figure}[htbp]
	\centering
	\begin{subfigure}{0.48\textwidth}
	\centering
	\includegraphics[width=\textwidth]{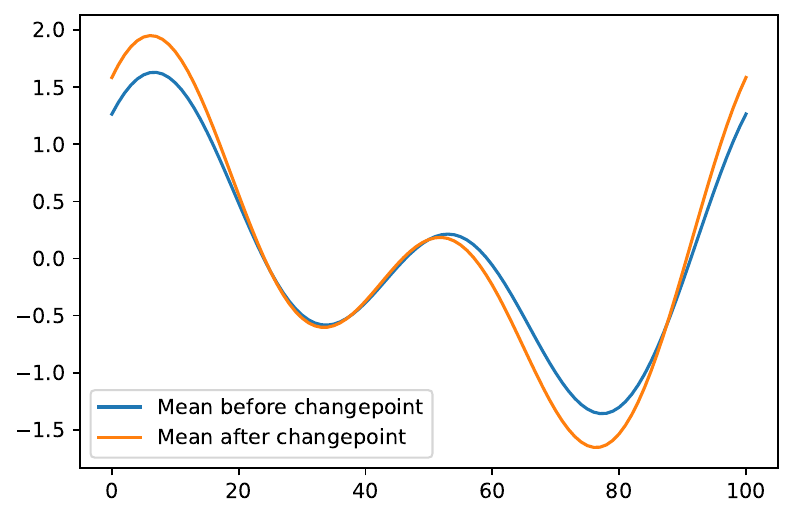}	
	\caption{}	
	\end{subfigure}
	\begin{subfigure}{0.48\textwidth}
	\centering
	\includegraphics[width=\textwidth]{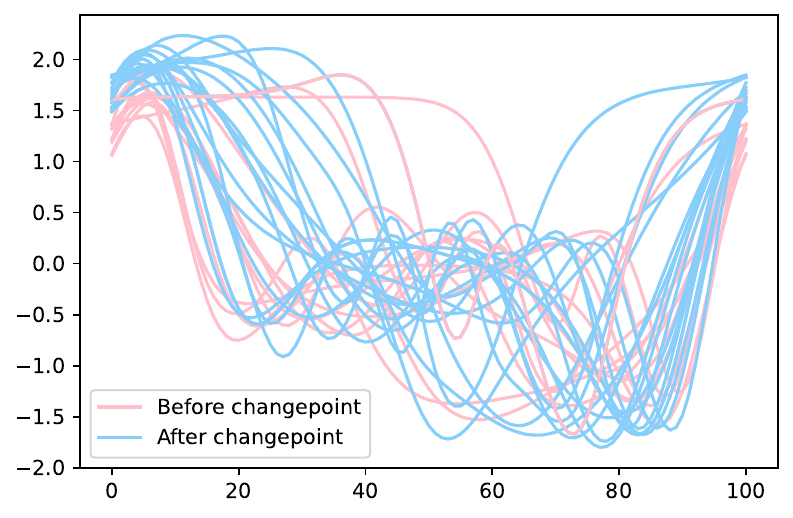}		
	\caption{}	
	\end{subfigure}
	\caption{Example of simulated data (a) Mean before changepoint $\mu$ and mean after changepoint $\mu + \delta$. (b) Simulated data, with color corresponding to whether the function was before or after the change. For this example, both elastic amplitude fully-functional and dimension-reduction-based approaches detected a changepoint with $\hat{k}^* = 12  = k^*$ and p-values of $0.000$. The cross-sectional approaches estimated $\hat{k}^*$ as $11$ (fully-functional) or $13$ (dimension-reduction) with p-values of $0.142$ and $0.319$, respectively. }
	\label{fig:sim_amp_ex}
\end{figure}

\noindent\textbf{Amplitude Changepoint}\\
\indent 
To discuss the basics of the simulation setup, we consider sample sizes of $n \in\{ 15, 30, 50, 75\}$, with an amplitude changepoint $k^* = 0.40 \cdot n$ so that $k^* \in \{6, 12, 20, 30\}$, respectively. 
For each simulation, the mean functions are generated as
\begin{align*}
\mu(t) &= a_0\cos(2\pi t) + b_0\sin(2\pi t) + a_1\cos(4\pi t) + b_1\sin(4\pi t) \\
\delta(t) &= \Delta\cos(2\pi t) +  \Delta\sin(2\pi t) + \Delta\cos(4\pi t) + \Delta\sin(4\pi t)
\end{align*}
where $a_i\sim U[0,1]$ and $b_i \sim U[0,1]$ for $i\in\{0,1\}$, and $\Delta$ is a signal strength parameter. 
The error term is taken as \begin{align*}
    \epsilon_i(t) &= Z_{1,i} + Z_{2,i} \frac{1}{\sqrt{2}}\cos(8\pi t) + Z_{3,i}\frac{1}{\sqrt{2}}\sin(8\pi t)
\end{align*}where $Z_{l,i} \overset{ind}{\sim} \mathcal{N}(0, \sigma^2_l)$, with $(\sigma_1, \sigma_2, \sigma_3) = (0.14, 0.10, 0.08)$. 
Finally, each function was warped with an independent, random warping function with variance $0.2$, generated similarly to Section 5.1 of \cite{harris2021elastic}. 
We consider results over 100 simulations for each value of $n$ and each signal strength $\Delta \in \{0, 0.04, 0.08, 0.12, 0.16\}$. 
In Figure \ref{fig:sim_amp_ex}, we plot an example of the simulated data with $n = 30$ and $\Delta = 0.16$, as well as its results using these test statistics. 
For each simulation, we compute score-based and fully-functional test statistics for both the cross-sectional \citep{aue2009estimation, aue:2018} and elastic amplitude changepoint methods. 

Overall changepoint detection results are presented for the dimension-reduction based methodology in Table \ref{tab:sim_power_pca}. The $\Delta = 0$ column (no change) represents the estimated Type I error of the detection procedure. 
We see that both approaches control Type I error well. 
As $\Delta$ and $n$ increase, it becomes easier to detect changepoints under the alternative hypothesis. 
The elastic approach uniformly improves power over the cross-sectional approach, detecting close to all changepoints when $\delta \geq 0.12$ and $n \geq 30$. 
In contrast, the cross-sectional approach only detects some changepoints and may not be consistent in this setting, as increasing $n$ does not always substantially improve the number of changepoints detected. 

\begin{table}[htbp]
    \centering
    \begin{tabular}{|c|c|c|c|c|c|}\hline
        Elastic PCA & $\Delta= 0$ & $\Delta = 0.04$ & $\Delta = 0.08$ & $\Delta = 0.12$ & $\Delta = 0.16$   \\ \hline \hline
        $n=15$ & $0.01$ & $0.07$ & $0.58$ & $0.77$ & $0.86$   \\ \hline
        $n=30$ & $0.02$ & $0.22$ & $0.92$ & $0.99$ & $1.00$   \\ \hline
        $n=50$ & $0.03$ & $0.25$ & $0.95$ & $1.00$ & $1.00$   \\ \hline
        $n=75$ & $0.01$ & $0.39$ & $0.98$ & $1.00$ & $1.00$   \\ \hline\hline
        Cross-sectional PCA & $\Delta= 0$ & $\Delta = 0.04$ & $\Delta = 0.08$ & $\Delta = 0.12$ & $\Delta = 0.16$   \\ \hline \hline
        $n=15$ & $0.01$ & $0.02$ & $0.03$ & $0.05$ & $0.07$   \\ \hline
        $n=30$ & $0.02$ & $0.09$ & $0.17$ & $0.18$ & $0.30$   \\ \hline
        $n=50$ & $0.02$ & $0.08$ & $0.13$ & $0.34$ & $0.36$   \\ \hline
        $n=75$ & $0.05$ & $0.04$ & $0.24$ & $0.24$ & $0.40$   \\ \hline
    \end{tabular}
    \caption{Proportion of simulations with estimated changepoint at the $\alpha = 0.05$ level for the amplitude change. (Top) Using the elastic PCA test statistic (Bottom) Using the cross-sectional PCA test statistic of \cite{aue2009estimation}.  }
    \label{tab:sim_power_pca}
\end{table}

We present respective results for the fully-functional approaches in Table \ref{tab:sim_power_ff}.
While Type I error is slightly inflated for both approaches, similar gaps in power between the elastic and cross-sectional approaches hold when the alternative is true, again suggesting that the cross-sectional approach will miss potential changes in the data. 
In general, the fully-functional test statistics tend to have more less power than the dimension-reduction-based test statistics.
\begin{table}[htbp]
    \centering
    \begin{tabular}{|c|c|c|c|c|c|}\hline
        Elastic fully-functional & $\Delta= 0$ & $\Delta = 0.04$ & $\Delta = 0.08$ & $\Delta = 0.12$ & $\Delta = 0.16$   \\ \hline \hline
        $n=15$ & $0.19$ & $0.18$ & $0.35$ & $0.54$ & $0.85$   \\ \hline
        $n=30$ & $0.07$ & $0.16$ & $0.27$ & $0.85$ & $1.00$   \\ \hline
        $n=50$ & $0.06$ & $0.11$ & $0.52$ & $1.00$ & $1.00$   \\ \hline
        $n=75$ & $0.09$ & $0.20$ & $0.86$ & $1.00$ & $1.00$   \\ \hline\hline
        Cross-sectional fully-functional & $\Delta= 0$ & $\Delta = 0.04$ & $\Delta = 0.08$ & $\Delta = 0.12$ & $\Delta = 0.16$   \\ \hline \hline
        $n=15$ & $0.13$ & $0.10$ & $0.13$ & $0.09$ & $0.15$   \\ \hline
        $n=30$ & $0.04$ & $0.12$ & $0.07$ & $0.16$ & $0.15$   \\ \hline
        $n=50$ & $0.06$ & $0.06$ & $0.12$ & $0.18$ & $0.20$   \\ \hline
        $n=75$ & $0.04$ & $0.08$ & $0.13$ & $0.21$ & $0.28$   \\ \hline
    \end{tabular}
    \caption{Proportion of simulations with estimated changepoint at the $\alpha = 0.05$ level for the amplitude change. (Top) Using the elastic fully-functional test statistic (Bottom) Using the cross-sectional fully-functional test statistic of \cite{aue:2018}.  }
    \label{tab:sim_power_ff}
\end{table}
We also compare the estimation of $k^*$ using a sample-size-normalized RMSE based on the estimated changepoints $\hat{k}_{mc}^*$ over each simulation $mc = 1, \dots, MC = 100$: $$\sqrt{\frac{1}{MC}\sum_{mc = 1}^{MC}\left(\frac{\hat{k}^*_{mc} - k^*}{n}\right)^2 },$$ with results in Table \ref{tab:sim_est_ff} for the fully-functional test statistics. 
The results mirror the testing results, with the elastic approach consistently estimating $k^*$ better compared to the cross-sectional approach. 
The cross-sectional approach only partially improves estimates as $\Delta$ increases or $n$ increases, again suggesting that cross-sectional approaches will be inadequate when there is phase variability in the data even if the sample size and signal strength are large.

\begin{table}[htbp]
    \centering
    \begin{tabular}{|c|c|c|c|c|c|}\hline
        Elastic fully-functional & $\Delta= 0$ & $\Delta = 0.04$ & $\Delta = 0.08$ & $\Delta = 0.12$ & $\Delta = 0.16$   \\ \hline \hline
        $n=15$ & $0.21$ & $0.21$ & $0.17$ & $0.10$ & $0.06$   \\ \hline
        $n=30$ & $0.23$ & $0.18$ & $0.10$ & $0.05$ & $0.03$   \\ \hline
        $n=50$ & $0.24$ & $0.16$ & $0.06$ & $0.01$ & $0.01$   \\ \hline
        $n=75$ & $0.23$ & $0.15$ & $0.05$ & $0.01$ & $0.00$   \\ \hline\hline
        Cross-sectional fully-functional & $\Delta= 0$ & $\Delta = 0.04$ & $\Delta = 0.08$ & $\Delta = 0.12$ & $\Delta = 0.16$   \\ \hline \hline
        $n=15$ & $0.19$ & $0.21$ & $0.21$ & $0.19$ & $0.18$   \\ \hline
        $n=30$ & $0.21$ & $0.21$ & $0.21$ & $0.18$ & $0.19$   \\ \hline
        $n=50$ & $0.22$ & $0.21$ & $0.18$ & $0.17$ & $0.14$   \\ \hline
        $n=75$ & $0.20$ & $0.21$ & $0.19$ & $0.14$ & $0.14$   \\ \hline
    \end{tabular}
    \caption{Sample-size-normalized RMSE of estimated change time for the amplitude change. (Top) Using the elastic fully-functional test statistic (Bottom) Using the cross-sectional fully-functional test statistic of \cite{aue:2018}.}
    \label{tab:sim_est_ff}
\end{table}

\vspace{.1in}\noindent\textbf{Phase Changepoint}\\
\indent To evaluate changepoint detection in the mean phase of functional data, We consider the same simulation setup with the exception of the following. 
We take $\delta(t)=0$, and the warping functions have mean $\gamma(t) = t$ before $k^*$.
After $k^*$, the mean of the warping functions is a randomly-generated warping function with variance $\Delta_{\gamma}$, where $\Delta_{\gamma} \in \{0, 0.075, 0.15, 0.225, 0.30\}$. 
The variance of the noise in each warping function was set to $0.05$. 

We give results in Tables \ref{tab:sim_power_phase_ff} and \ref{tab:sim_est_phase_ff}. 
The elastic test for a phase change and the cross-sectional test perform similarly for both testing and changepoint estimation. 
The cross-sectional approach has a bit higher Type I error and slightly more power when $n = 15$. 
The results are essentially same between the two approaches for the estimation problem.

\begin{table}[htbp]
    \centering
    \begin{tabular}{|c|c|c|c|c|c|}\hline
        Elastic fully-functional & $\Delta= 0$ & $\Delta = 0.04$ & $\Delta = 0.08$ & $\Delta = 0.12$ & $\Delta = 0.16$   \\ \hline \hline
        $n=15$ & $0.06$ & $0.41$ & $0.81$ & $0.81$ & $0.93$   \\ \hline
        $n=30$ & $0.03$ & $0.61$ & $0.82$ & $0.95$ & $0.97$   \\ \hline
        $n=50$ & $0.04$ & $0.75$ & $0.95$ & $0.95$ & $0.98$   \\ \hline
        $n=75$ & $0.00$ & $0.88$ & $0.92$ & $0.97$ & $1.00$   \\ \hline\hline
        Cross-sectional fully-functional & $\Delta= 0$ & $\Delta = 0.04$ & $\Delta = 0.08$ & $\Delta = 0.12$ & $\Delta = 0.16$   \\ \hline \hline
        $n=15$ & $0.14$ & $0.65$ & $0.88$ & $0.93$ & $0.95$   \\ \hline
        $n=30$ & $0.07$ & $0.65$ & $0.85$ & $0.94$ & $0.99$   \\ \hline
        $n=50$ & $0.08$ & $0.76$ & $0.96$ & $0.92$ & $0.98$   \\ \hline
        $n=75$ & $0.08$ & $0.88$ & $0.93$ & $0.97$ & $1.00$   \\ \hline
    \end{tabular}
    \caption{Proportion of simulations with estimated changepoint at the $\alpha = 0.05$ level for the phase change. (Top) Using the elastic fully-functional test statistic (Bottom) Using the cross-sectional fully-functional test statistic of \cite{aue:2018}.  }
    \label{tab:sim_power_phase_ff}
\end{table}

\begin{table}[htbp]
    \centering
    \begin{tabular}{|c|c|c|c|c|c|}\hline
        Elastic fully-functional & $\Delta= 0$ & $\Delta = 0.04$ & $\Delta = 0.08$ & $\Delta = 0.12$ & $\Delta = 0.16$   \\ \hline \hline
        $n=15$ & $0.21$ & $0.15$ & $0.04$ & $0.05$ & $0.05$   \\ \hline
        $n=30$ & $0.21$ & $0.12$ & $0.05$ & $0.04$ & $0.04$   \\ \hline
        $n=50$ & $0.19$ & $0.11$ & $0.04$ & $0.04$ & $0.01$   \\ \hline
        $n=75$ & $0.21$ & $0.04$ & $0.05$ & $0.04$ & $0.00$   \\ \hline\hline
        Cross-sectional fully-functional & $\Delta= 0$ & $\Delta = 0.04$ & $\Delta = 0.08$ & $\Delta = 0.12$ & $\Delta = 0.16$   \\ \hline \hline
        $n=15$ & $0.22$ & $0.15$ & $0.07$ & $0.07$ & $0.03$   \\ \hline
        $n=30$ & $0.22$ & $0.12$ & $0.06$ & $0.04$ & $0.02$   \\ \hline
        $n=50$ & $0.22$ & $0.11$ & $0.05$ & $0.04$ & $0.01$   \\ \hline
        $n=75$ & $0.22$ & $0.05$ & $0.05$ & $0.04$ & $0.01$   \\ \hline
    \end{tabular}
    \caption{Sample-size-normalized RMSE of estimated change time for the phase change. (Top) Using the elastic fully-functional test statistic (Bottom) Using the cross-sectional fully-functional test statistic of \cite{aue:2018}.}
    \label{tab:sim_est_phase_ff}
\end{table}

\begin{figure}[htbp]
	\centering
	\begin{subfigure}{0.32\textwidth}
	\centering
	\includegraphics[width=\textwidth]{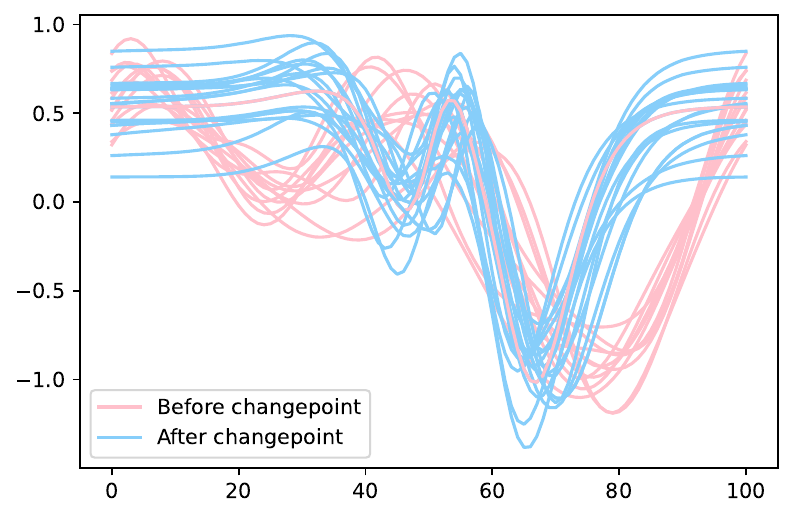}	
	\caption{}	
	\end{subfigure}
	\begin{subfigure}{0.32\textwidth}
	\centering
	\includegraphics[width=\textwidth]{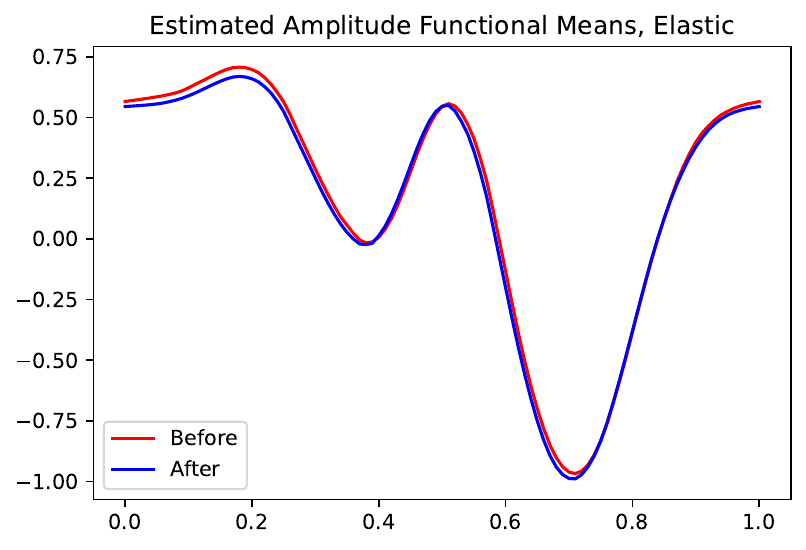}		
	\caption{}	
	\end{subfigure}
	\hfill
	\begin{subfigure}{0.32\textwidth}
	\centering
	\includegraphics[width=\textwidth]{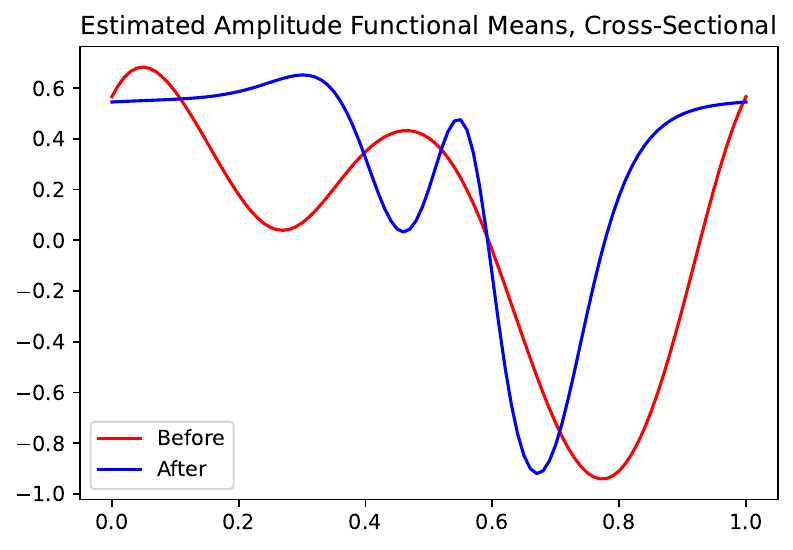}		
	\caption{}	
	\end{subfigure}

	\caption{A phase simulation example with $\Delta_\gamma = 0.3$ and $n = 30$. (a) Simulated functional data (b) Estimated amplitude means using the elastic fully-functional test statistic. (c) Estimated means using the cross-sectional fully-functional test statistic. Both approaches detected the correct changepoint $k^* = 12$ with a p-value of $0.000$.}
	\label{fig:phase_example}
\end{figure}

While the cross sectional approach is able to detect changes, it will not be able to identify the data-generating processes, as it interprets all variability in the data as amplitude variability. 
To demonstrate this, we plot an example in Figure \ref{fig:phase_example}. 
While the mean amplitude functions are the same before and after the changepoint, the data warping in (a) shows considerably different observed functions. 
The estimated means based on the elastic and cross-sectional fully-functional statistics are shown in (b) and (c) respectively.
While the elastic approach estimates nearly the same amplitude function before and after the changepoint, phase variability is present and the cross-sectional detection procedure does not recognize the similarity between the mean amplitude functions of the data. 

\subsection{MERRA-2 stratospheric temperature}\label{sec:MERRA2}

We now present our data analysis on stratospheric temperature of the climate reanalysis data MERRA-2 \citep{MERRA2}. 
Using data from the years 1984-1998, we aim to evaluate changes related to the eruption of Mt.\ Pinatubo in June 1991. 
See, for example, \cite{thompson_2009} for information on variability and variables related to Mt.\ Pinatubo's eruption. 
For this study, we focus on daily stratospheric temperature near the 50 millibar pressure surface.
We run each detection procedure at 3,312 different locations on a grid.
Since we are primarily interested in an amplitude changepoint, we do not present results based on a change in the phase function.

\begin{figure}[htbp]
	\centering
	\begin{subfigure}{0.35\textwidth}
	\centering
	\includegraphics[width=\textwidth]{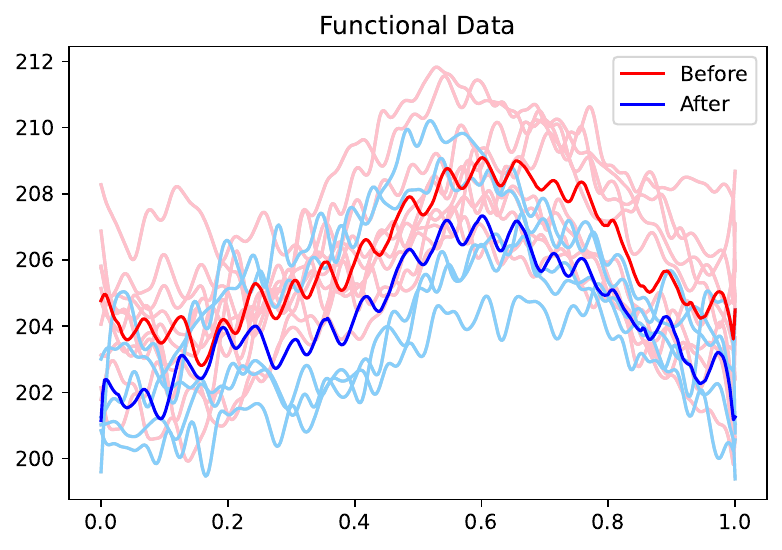}	
	\caption{}	
	\end{subfigure}
	\begin{subfigure}{0.35\textwidth}
	\centering
	\includegraphics[width=\textwidth]{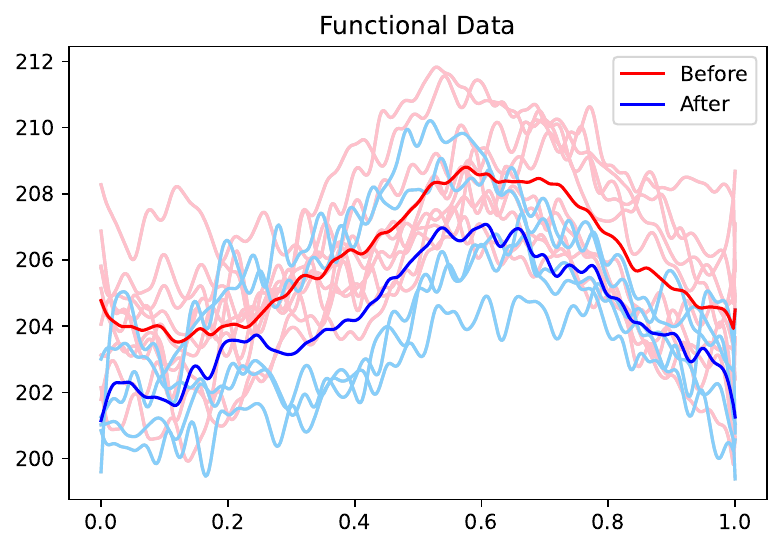}		
	\caption{}	
	\end{subfigure}
	\begin{subfigure}{0.26\textwidth}
	\centering
	\includegraphics[width=\textwidth]{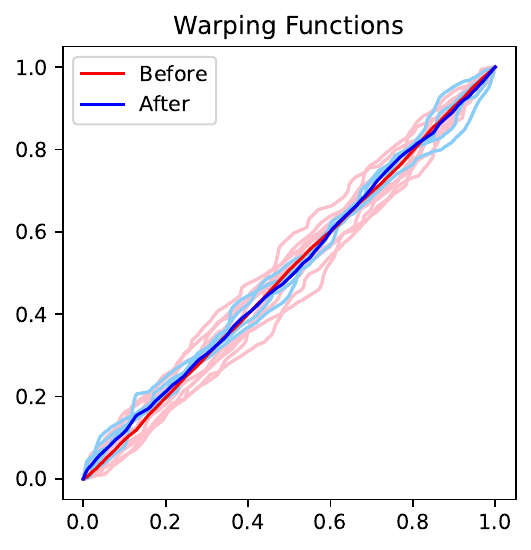}		
	\caption{}	
	\end{subfigure}
	\hfill
	\begin{subfigure}{0.35\textwidth}
	\centering
	\includegraphics[width=\textwidth]{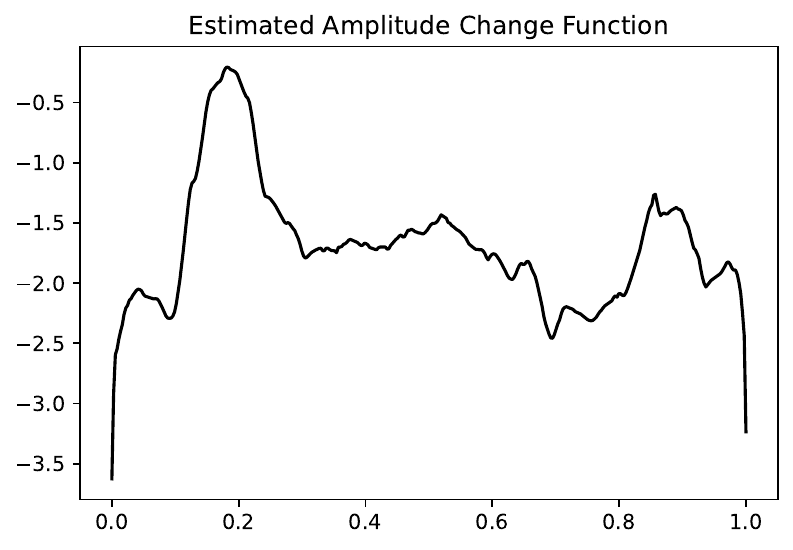}		
	\caption{}	
	\end{subfigure}
	\begin{subfigure}{0.35\textwidth}
	\centering
	\includegraphics[width=\textwidth]{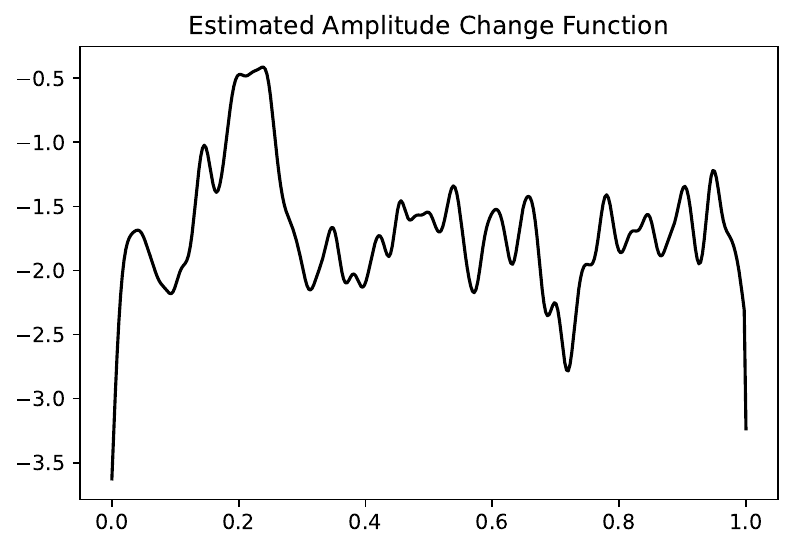}		
	\caption{}	
	\end{subfigure}
	\hspace{.26  \textwidth}

	\caption{An example of MERRA-2 stratospheric temperature (in Kelvin) fully-functional results at $0^\circ$W and $2^\circ$N. The elastic approach had $\hat{k}^* = 1993$ with a p-value of $0.028$, and the cross-sectional had $\hat{k}^* = 1993$ with a p-value of $0.067$. (a) Functional data and estimated means after alignment. (b) Functional data and estimated means before alignment. (c) Estimated warping functions. (d) Estimated change for elastic approach. (e)  Estimated change for cross-sectional approach. (e) Warping functions.}
	\label{fig:merra2_ff_example}
\end{figure}

We begin by presenting fully-functional results for a single location in Figure \ref{fig:merra2_ff_example}. 
In (a) and (b), we plot the original functional data along with the estimated mean functions for the elastic approach (a) and the cross-sectional approach (b) before and after the estimated change. 
The elastic approach appears to align weather patterns, maintaining cyclical behavior in the temperature throughout the year. 
In contrast, the cross-sectional approach averages these over years while ignoring phase variability. 
The distinction between these two estimates is filtered into estimates of the change function in (d) and (e).
The estimated change function from the elastic approach (d) does not have major oscillations since the cyclical behavior is shared between the two mean functions. 
For the cross-sectional approach of \cite{aue:2018}, however, this variability is propagated into the estimate of the change function, resulting in an estimate that is more noisy. 
By accounting for phase variability in the original data, the estimate of the change is more easily isolated and estimated. 
In panel (c), we plot the warping functions, colored based on if they were before or after the changepoint. 
While some phase variability is present and substantially affects the estimates, functions do not have to be warped drastically, and there does not appear to be a trend or change in the warping functions.

\begin{figure}[htbp]
	\centering
	\begin{subfigure}{0.98\textwidth}
	\centering
	\includegraphics[width=\textwidth]{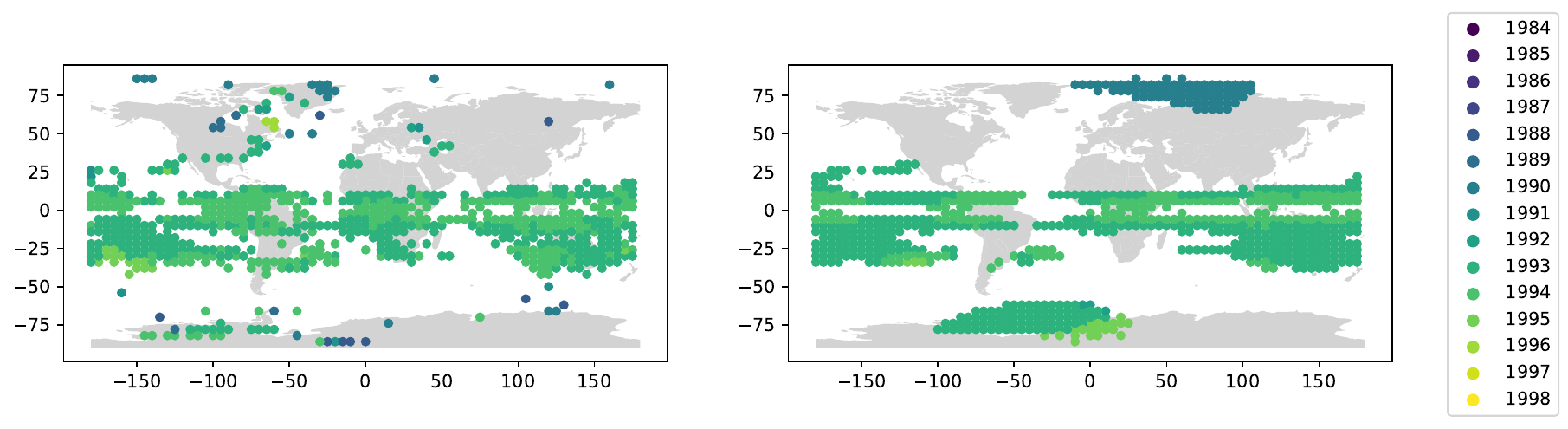}	
	\caption{}	
	\end{subfigure}
	\hfill
	\begin{subfigure}{0.98\textwidth}
	\centering
	\includegraphics[width=\textwidth]{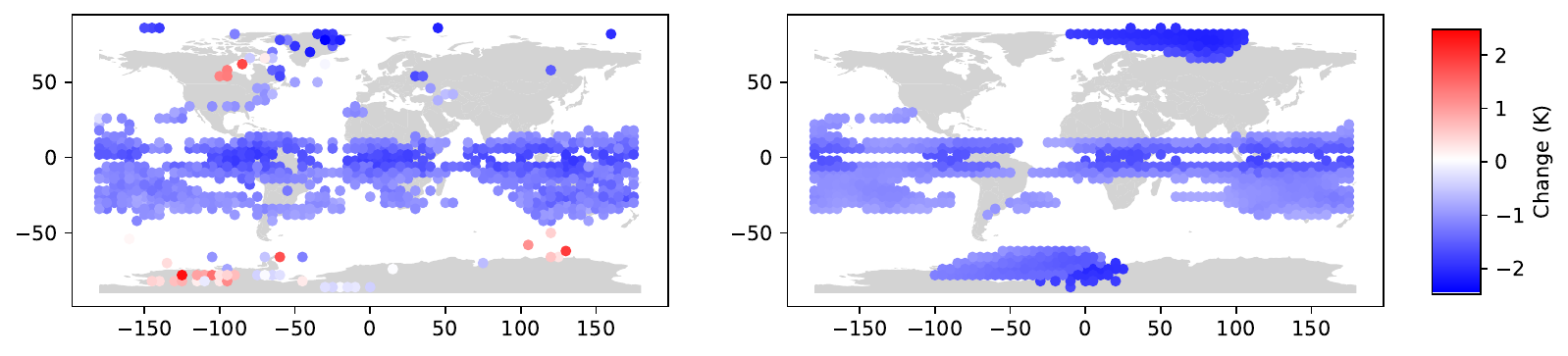}		
	\caption{}	
	\end{subfigure}
	\caption{MERRA-2 stratospheric temperature fully-functional results. (a) Estimated change years $\hat{k}^*$ with p-values less than $0.05$ using elastic approach (Left) and cross-sectional approach (Right). (b) Averaged estimated change $\int_0^1 \hat{\delta}(t) dt$ with p-values less than $0.05$ for elastic approach (Left) and cross-sectional approach (Right).}
	\label{fig:merra2_ff}
\end{figure}

\begin{figure}[htbp]
	\centering
	\begin{subfigure}{0.98\textwidth}
	\centering
	\includegraphics[width=\textwidth]{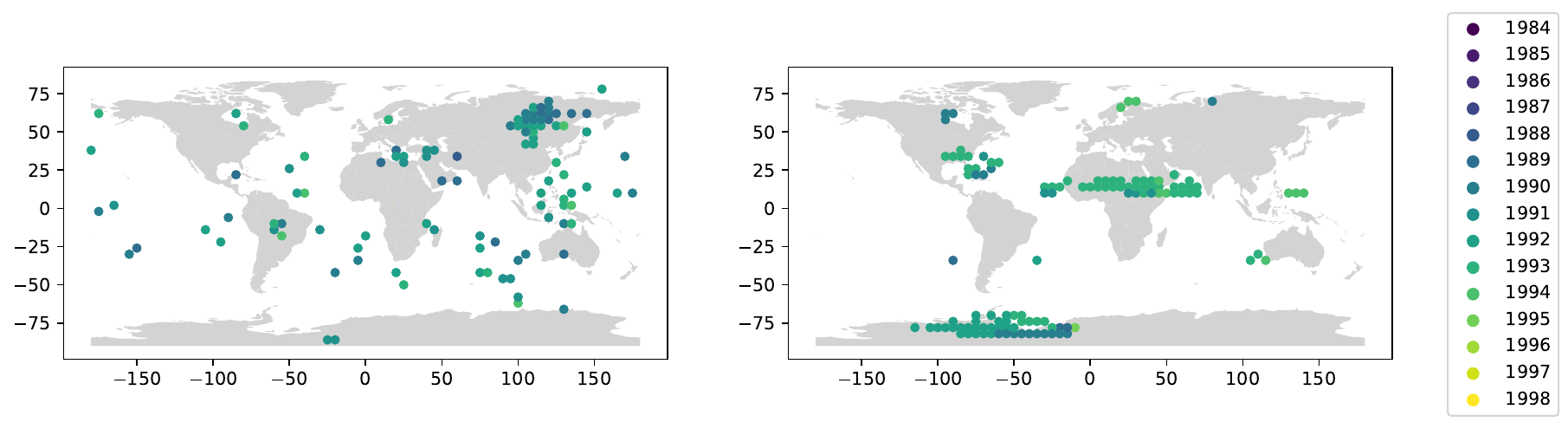}	
	\caption{}	
	\end{subfigure}
	\hfill
	\begin{subfigure}{0.98\textwidth}
	\centering
	\includegraphics[width=\textwidth]{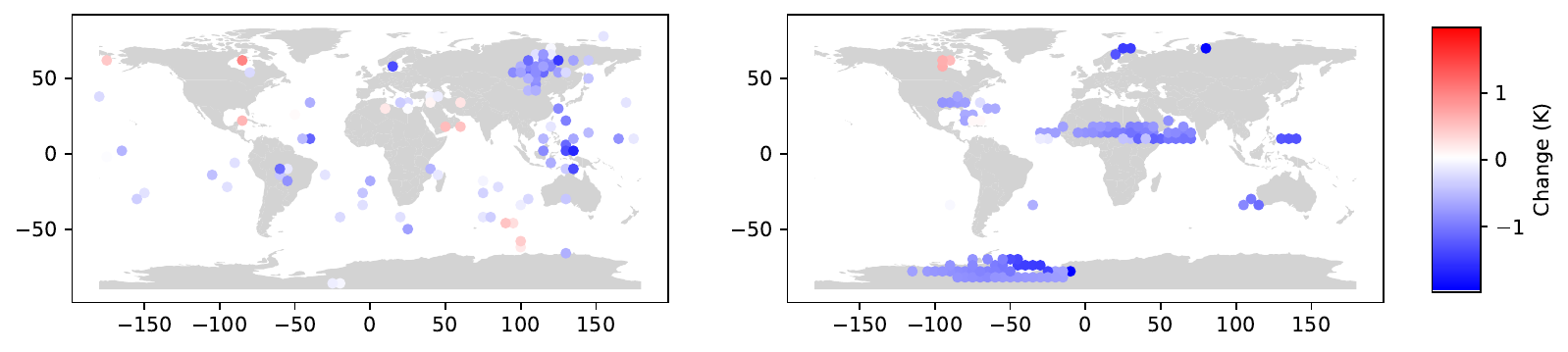}		
	\caption{}	
	\end{subfigure}
	\caption{MERRA-2 stratospheric score-based results. (a) Estimated change years $\hat{k}^*$ with p-values less than $0.05$ using elastic approach (Left) and cross-sectional approach (Right). (b) Averaged estimated change $\int_0^1 \hat{\delta}(t) dt$ with p-values less than $0.05$ for elastic approach (Left) and cross-sectional approach (Right).}
	\label{fig:merra2_score}
\end{figure}

We present the results for all locations in Figure \ref{fig:merra2_ff}. 
Locations were only plotted if there was an associated p-value less than $0.05$. 
For the most part, the elastic and cross-sectional approaches have similar results.
Both detect changes in the tropics in 1993 and 1994, where the impact of Mt.\ Pinatubo is particularly strong. 
The elastic approach detects fewer changes in the poles, which are less likely to be associated with the eruption of Mt.\ Pinatubo.
To interpret the changes detected, we see that the changes in the tropics correspond to a decrease in temperature. 
Combined with the time of the changepoint, both methodologies detect when the stratospheric temperature ``returns to normal'' after a period of increase in 1991 and 1992 immediately following the eruption. 
The estimated changes are comparable between the two approaches. We summarize the detected changepoints in Table \ref{tab:merra2_ff_summary}. 
The vast majority of the detected changepoints had $\hat{k}^* = 1992$ or $\hat{k}^* = 1993$, with an associated decrease in temperature on average. 
In Figure \ref{fig:merra2_score}, we plot results based on the dimension reduction test statistics. 
Although fewer changepoints were detected, overall the pattern and nature of the changes are similar to those of the fully-functional test statistics.

\begin{table}[htbp]
    \centering
    \begin{tabular}{c|c|c||c|c|c|}\hline
        Year & \# of detections & Avg.\ change (K) & Year & \# of detections & Avg.\ change (K)  \\ \hline
        1984 & 0 & N/A & 1992 & 350 & -1.12 \\  \hline
        1985 & 0 & N/A & 1993 & 291 & -1.41\\  \hline
        1986 & 0 & N/A & 1994 & 17 & -1.18\\  \hline
        1987 & 9 & 0.03 & 1995 & 3 &  -1.64 \\  \hline
        1988 & 6 & 1.60 & 1996 & 0 & N/A \\  \hline
        1989 & 19 & -1.36 & 1997 & 0 & N/A\\  \hline
        1990 & 6 & -0.35 & 1998 & 0 & N/A\\  \hline
        1991 & 6 & -0.97 & & &\\  \hline
    \end{tabular}
    \caption{Summary of detected test statistics by the elastic fully-functional method. }
    \label{tab:merra2_ff_summary}
\end{table}

\begin{figure}
    \centering
    	\begin{subfigure}{0.98\textwidth}
	\centering
	\includegraphics[width=\textwidth]{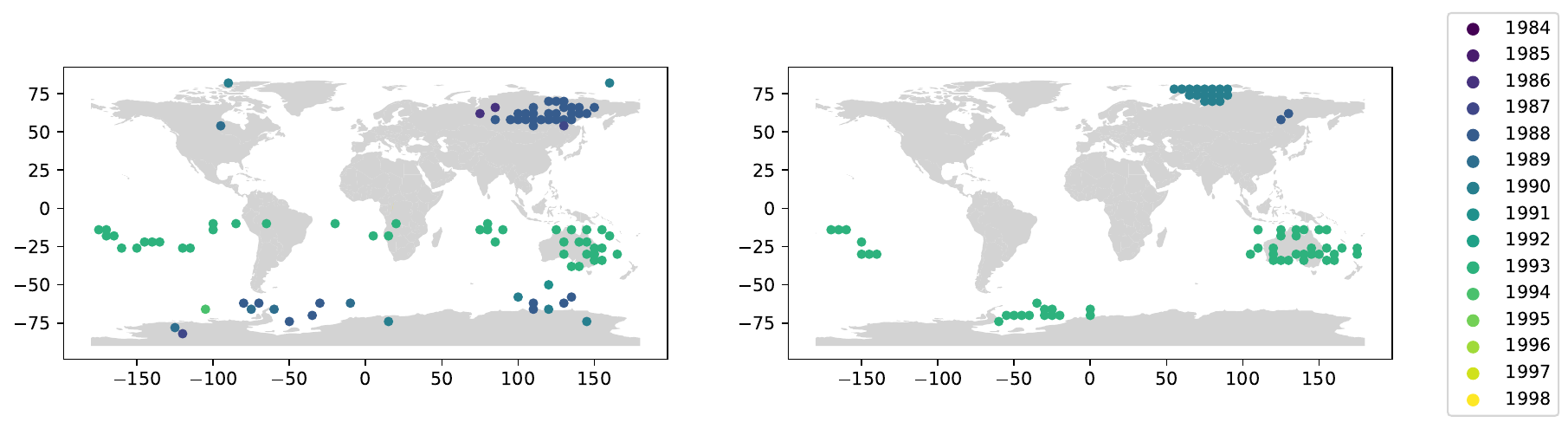}	
	\caption{}	
	\end{subfigure}
		\begin{subfigure}{0.98\textwidth}
	\centering
	\includegraphics[width=\textwidth]{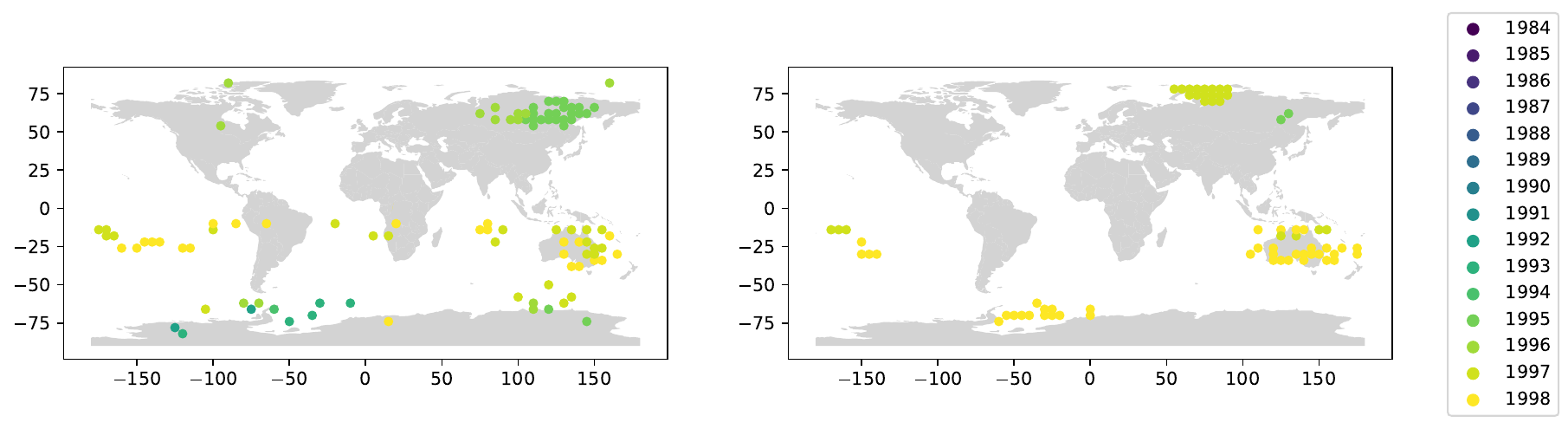}	
	\caption{}	
	\end{subfigure}
		\begin{subfigure}{0.98\textwidth}
	\centering
	\includegraphics[width=\textwidth]{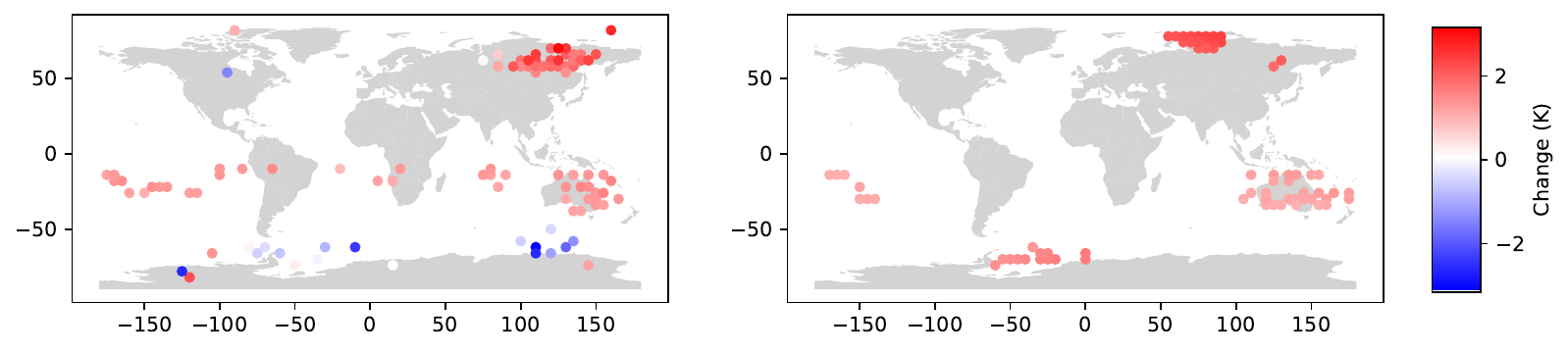}	
	\caption{}	
	\end{subfigure}
    \caption{Fully-functional epidemic results. (a) Estimated change years $\hat{k}^*_1$ with p-values less than $0.05$ using elastic approach (Left) and cross-sectional approach (Right) (b) The same as (a) with $\hat{k}^*_2$ (c) Averaged estimated change $\int_0^1 \hat{\delta}(t) dt$ for locations with p-values less than $0.05$. }
    \label{fig:MERRA2_epidemic}
\end{figure}

In our analysis, we appear to primarily detect a decrease in temperature after the year 1992 or so. 
Since Mt.\ Pinatubo is expected to temporarily increase stratospheric temperatures immediately after the eruption, we appear to be detecting when the temperature ``returns to normal'' as suggested before.
To more directly target the eruption, we consider the epidemic-type changepoint model studied in \cite{aston_2012}: 
\begin{align*}
    f_i = \mu + \delta \vmathbb{1}(k_1^* < i \leq k_2^*) + \epsilon_i
\end{align*}where the only difference with the ``at-most-one-change'' model $f_i = \mu + \delta \vmathbb{1}(i > k^*) + \epsilon_i$ is the indicator $ \vmathbb{1}(k_1^* < i \leq k_2^*)$. 
Fully-functional and dimension-reduction elastic approaches for detecting this type of changepoint and for estimating $k_1^*$ (when the change period begins) and $k_2^*$ (when the change period ends) follow naturally from the test statistics and results in \cite{aston_2012} and Section \ref{sec:efchange}. 
We present results for the epidemic model in Figure \ref{fig:MERRA2_epidemic}. 
While fewer changepoints are detected compared to the at-most-one-change model, the beginning of the changepoints $k^*_1$ are estimated earlier in the tropics, and these changes are associated with an increase in temperature rather than a decrease.

\begin{figure}
    \centering
	\includegraphics[width=\textwidth]{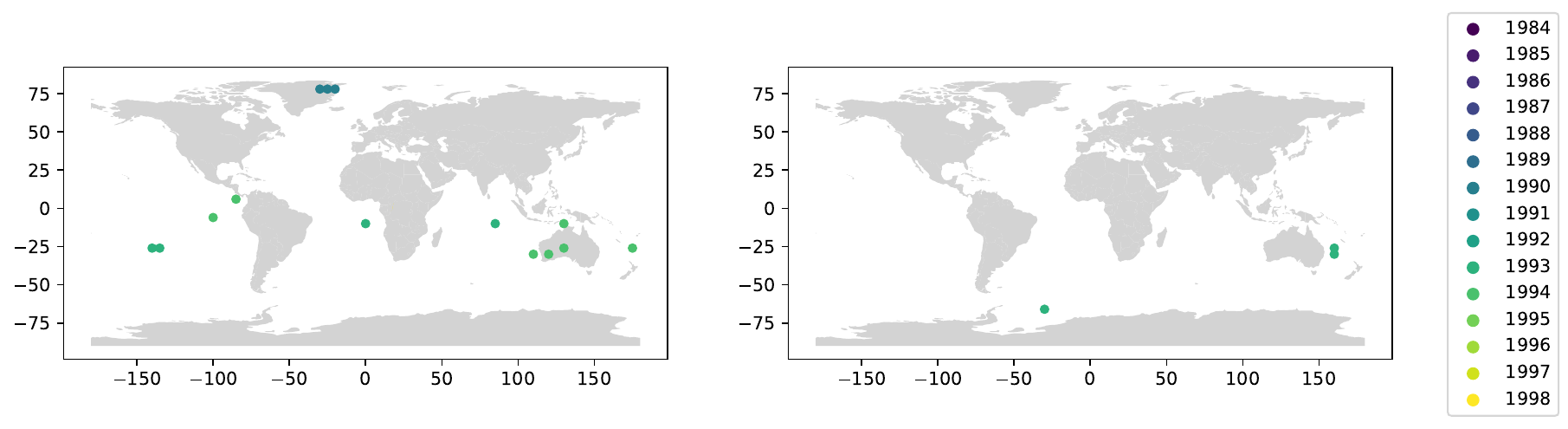}	

    \caption{Fully-functional results from the at-most-one-change model with a Benjamini-Hochberg p-value correction. Estimated change years $\hat{k}^*$ with Benjamini-Hochberg p-values less than $0.05$ using elastic approach (Left) and cross-sectional approach (Right).}
    \label{fig:MERRA2_bh}
\end{figure}

Finally, we evaluate the sensitivity of the detected changepoints to multiple testing correction. 
In Figure \ref{fig:MERRA2_bh}, we plot results with Benjamini-Hochberg corrected p-values for the fully-functional test statistics analogous to Figure \ref{fig:merra2_ff}.
We see that very few changepoints are detected, but the elastic approach detects substantially more changepoints after p-value adjustment. 

\section{Conclusion}
\label{sec:conclusion}
We have proposed a new flexible elastic approach to functional changepoint detection that handles phase and amplitude variability in the data accordingly. 
We have demonstrated its advantages over the state-of-the-art methods in the functional changepoint literature using both simulated and real datasets. 
Unlike the current cross-sectional methods, the elastic method accurately estimates the mean of the underlying data-generating mechanism and also handles phase as a nuisance or as information suggesting a potential changepoint in the functional data.
Correctly estimating the underlying mean before and after the change can have large impacts on interpretability of the data and implementation of downstream analysis. 

\edit{In a simulation study, the procedures correctly detect a changepoint and estimate its time when there is either an amplitude or phase changepoint in the functional time series.
In contrast, cross-sectional approaches miss detecting changepoints when there is a change in the mean amplitude, and the time of changepoint is not estimated well. 
In an analysis of stratospheric temperature, both the cross-sectional approaches and our introduced test statistics detect changes is the tropics after the eruption of Mt.\ Pinatubo. 
However, the test statistics that take into account phase variability result in a less noisy estimate of the change function, increasing its interpretability and smoothness. }

For future work, we look to extend the method to the multiple change-point problem. 
This extension will not be trivial as scalability is a problem as shown in \cite{harris:21}.
Multiple changepoints can be detected in $S_{n,k}$ if using a multiple peak detection scheme, but determining the $p$-value is not straightforward, and the test statistics will have to be defined carefully.
\edit{Another interesting question is the performance of the test statistics in the presence of outliers.
For example, an analysis of aerosol optical depth, which immediately increases after a volcanic eruption and returns to lower levels relatively quickly, would be illustrative.
Other work has proposed robustified fully-functional statistics \citep{wegner2022robust}, and application of their test statistics to our setting would be relatively straightforward. 
In addition, while we have treated each location individually, comprehensively studying the spatial or multivariate problem of how functional time series at different locations relate to each other would be a fruitful direction for future research that the authors are currently studying.
\cite{paynabar2016change}, \cite{wang:2022}, and \cite{moradi_2023} approach this problem in three different ways. 
Such an approach will also lend itself more to the multiple testing problem. 
}
Additionally, we can extend this from curves to trajectories that lie on Riemannian manifolds $\mathcal{M}$.
In this case, one has to account for the non-zero curvature of the space and in particular, the calculation of the gradient in the Fisher Rao metric and computation of the required statistics. 

\section*{Acknowledgment}
This paper describes objective technical results and analysis. Any subjective views or opinions that might be expressed in the paper do not necessarily represent the views of the U.S. Department of Energy or the United States Government. This work was supported by the Laboratory Directed Research and Development program at Sandia National Laboratories; a multi-mission laboratory managed and operated by National Technology and Engineering Solutions of Sandia, LLC, a wholly owned subsidiary of Honeywell International, Inc., for the U.S. Department of Energy's National Nuclear Security Administration under contract DE-NA0003525.

\baselineskip=20pt
\bibliography{reference}

\end{document}